\documentclass[preprint]{aastex}
\usepackage{graphicx}

\shorttitle{X-ray reflected spectra from accretion disk models}
\shortauthors{Garc\'{\i}a \& Kallman}

\begin{document}

\title{X-ray reflected spectra from accretion disk models. \\
I. Constant density atmospheres}

\author{J. Garc\'ia}
\affil{The Catholic University of America, IACS, Physics Department, Washington DC 20064; and
NASA Goddard Space Flight Center, Greenbelt, MD 20771}
\email{javier@milkyway.gsfc.nasa.gov}

\author{T.R. Kallman}
\affil{NASA Goddard Space Flight Center, Greenbelt, MD 20771}
\email{timothy.r.kallman@nasa.gov}

%
\begin{abstract}
We present new models for illuminated accretion disks, their structure 
and reprocessed emission. We consider the effects of incident X-rays on the 
surface of an accretion disk by solving simultaneously the equations of radiative
transfer, energy balance and ionization equilibrium over a large range of
column densities. We assume plane-parallel geometry and azimuthal
symmetry, such that each calculation corresponds to a ring at a given
distance from the  central object. Our models include recent and complete
atomic data for K-shell processes of the iron and oxygen isonuclear sequences.
We examine the effect on the spectrum of fluorescent K$\alpha$ line emission 
and absorption in the emitted spectrum. We also explore the dependence of the 
spectrum on the strength of the incident X-rays and other input parameters, 
and discuss the importance of Comptonization on the emitted spectrum. 
\end{abstract}
%
%
\section{Introduction}
The X-ray spectra from active galactic nuclei (AGN) and X-ray binaries often 
show evidence of interaction between radiation emitted near the compact object and
the nearby gas, which leads to signatures imprinted on the observed
spectrum. The effects of this reflection include iron K line emission, 
in the range 6-8 keV, which is observed from nearly all accreting compact 
sources \citep{got95}; and the flattening of the spectrum above 10 keV, 
usually called high energy bump or the Compton hump, since it originates due to
the Compton scattering of photons by cold electrons in the gas.

The profile of the iron K line carries important information about the physics
of the material around the compact object. Since the emission may 
occur in a relatively small region close to the center, the line profile can be
affected by relativistic effects. The best-known example is the Seyfert 1 
galaxy MCG-6-30-15 \citep{tan95, iwa99}, for which the iron line appears 
to be broad and skewed well beyond the instrumental resolution. Model line
profiles for Schwarzschild and Kerr metrics have been calculated 
by \cite{fab89} and \cite{lao91}, respectively. \cite{bre06} have used this 
information to constrain the spin of the black hole harbored by
this AGN.

Observations of other accreting systems such as low mass X-ray binaries (LMXB)
have also revealed the iron K line in emission. \cite{bha07} reported the detection 
of a broad iron line emission in the {\it XMM Newton} spectrum of the LMXB system 
Serpens X-1. Using {\it Suzaku} observations, \cite{cac08} confirmed this detection
and found similar emission in two other LMXBs (4U 1820-30 and GX 349+2). Assuming relativistic 
broadening and applying the same models used in AGN observations, these authors
have derived estimates for the inner radius of the accretion disk which can be used as an
upper limit for the radius of the neutron star. Similar behavior appears
in the millisecond pulsar SAX J1808.4-3658 \citep{cac09}. Since the temperature 
of the surface layers of an accretion disk around a stellar-mass
black hole or a neutron star is expected to be much higher than the 
temperature in the accretion disk of an AGN, the Compton down scattering 
of photons can also modify the emission profile and even contribute to the 
broadening of the iron K line \citep{ros07}. \cite{rei09} found evidence for such a 
situation in the {\it Suzaku} spectrum of the LMXB 4U 1705-44.

An X-ray line propagating in a dense gas has a non-negligible probability 
of interacting with an electron, which leads to a down scattering of photons. 
The maximum energy shift per scattering of a photon with energy $E_c$ due to the electron recoil 
(for backscattered photons at 180$^{\circ}$), has magnitude $\Delta E_{max} = 2 E_c^2/(m_ec^2+2E_c)$, 
where $m_ec^2$ is the electron rest-mass energy. Many scatterings will
therefore produce a discernible ``Compton shoulder" between $E_c$ and 
$E_c - \Delta E_{max}$ \citep{poz79,ill79,mat91}. 
Although the Compton shoulder and the iron K${\beta}$ emission line are weaker
than the iron K$\alpha$ emission line, both have 
been detected in AGN observations, especially those from the X-ray
observatory {\it Suzaku} \citep{mar07,ree07,yaq07}, and at least in one LMXB,
GX 301-2 \citep{wat03}.

Initial studies of reprocessed radiation by cold matter were due to \cite{lig80} and
\cite{lig81}, who derived the Green's functions for the scattering of photons by cold
electrons and discussed the implications for AGN observations \citep{lig88}. \cite{geo91}
included line production in their Monte Carlo calculations, while photoionization
equilibrium was included by \cite{zyc94}. Several authors expanded these studies
assuming constant density along the vertical direction of the disk 
\citep{don92,ros93,mat93,cze94,kro94,mag95,ros96,mat96,pou96,bla99,ros05}, while hydrostatic
calculations have been carried out by \cite{roz96,nay00, nay01, bal01, dum02} and \cite{ros07}.

Nevertheless, one of the most significant limitations inherent in current reflection
models is the limited treatment of the physics governing the atomic processes affecting 
the excitation and emission from the ions in the atmosphere, especially the iron K-shell
structure. In most cases, reflection models implement analytic fits to the partial 
Hartree-Dirac-Slater photoionization cross sections of \cite{ver95}, together with the line
energies, fluorescence and Auger yields from \cite{kas93}, which have been proven
to be over simplified and incomplete \citep{gor05}. In some cases, the models lack atomic data
for any of the neutral species (except for H and He).

In this paper, we present new models for illuminated accretion disks and their 
structure implementing state-of-the-art atomic data for the isonuclear
sequences of iron \citep{kal04}, and oxygen \citep{gar05}. In addition, the energy, spatial
and angular resolution of our calculations are greater than previous works.
The ionization balance calculations are performed in detail using the 
latest version of the photoionization code {\sc xstar}. The radiation transfer equation
is solved at each depth, energy and angle using the Feautrier formalism. We assume
plane-parallel geometry and azimuthal symmetry. For simplicity in this paper we assume 
constant density along the vertical direction of the atmosphere, and we leave the
hydrostatic calculation for future publication.

This paper is organized in the following way. In \S~\ref{secmet} we describe the 
theory and numerical methods used to solve radiation transfer, ionization and energy
equilibrium, and the atomic data. In \S~\ref{secres} we present a set of
models for constant density atmospheres at different degrees of ionization, viewing
and incidence angles, as well as the effect of the abundance on the reflected spectrum.
Finally, the main conclusions are summarized in \S~\ref{seccon}.
%
%
\section{Methodology}\label{secmet}
In this section we describe the theory used in our calculations, in particular, 
the numerical methods to solve the radiation transfer equation, the ionization
equilibrium, energy conservation and the new atomic data incorporated in our
models.
%
%
\subsection{Radiative Transfer}\label{secrad}
The standard form of the transfer equation for a one dimensional, plane-parallel
atmosphere is given by
\begin{equation}\label{eqrtz}
\mu \frac{\partial I(z,\mu,E)}{\partial z} = \eta(z,E) - \chi(z,E)I(z,\mu,E),
\end{equation}
where $\mu$ is the cosine of the angle with respect to the normal, and $\eta(z,E)$
and $\chi(z,E)$ are the total emissivity and opacity, respectively. It is convenient
to write this equation in terms of the Thomson optical depth, $d\tau \equiv -\alpha_T dz =
- \sigma_{T} n_e dz$, where $\sigma_{T}$ is the Thomson cross section ($=6.65\times 10^{-25}$ cm$^2$),
and $n_e$ is the electron number density, such that
\begin{equation}\label{eqrtt}
\omega(\tau,E)\mu\frac{\partial I(\tau,\mu,E)}{\partial\tau} =
I(\tau,\mu,E) - \frac{\eta(\tau,E)}{\chi(\tau,E)}
\end{equation}
where we have defined 
\begin{equation}
\omega(\tau,E) \equiv \frac{\alpha_T}{\chi(\tau,E)}.
\end{equation}
To solve the equation of radiative transfer we use Lambda iteration in the 
Feautrier formalism, as described in \S~6.3 of \cite{mih78}. 
By restricting $\mu$ to the half-range $0\leq\mu\leq 1$ to express 
the radiation field in its incoming and outgoing
components, and using the symmetric and antisymmetric averages:
\begin{equation}\label{equav}
u(\tau,\mu,E)=\frac{1}{2}[I(\tau,+\mu,E)+I(\tau,-\mu,E)]
\end{equation}
\begin{equation}\label{eqvav}
v(\tau,\mu,E)=\frac{1}{2}[I(\tau,+\mu,E)-I(\tau,-\mu,E)],
\end{equation}
one can rewrite the Equation~(\ref{eqrtt}) as a second order differential equation
\begin{equation}\label{eqrt}
\mu^2\omega^2(E)\frac{\partial^2 u(\mu,E)}{\partial\tau^2} +
\mu^2\omega(E)\frac{\partial\omega(E)}{\partial\tau}
\frac{\partial u(\mu,E)}{\partial\tau} = u(\mu,E) - S(E)
\end{equation}
where we have omitted the obvious dependence on $\tau$. Notice 
that the second term on the left hand side of the equation appears since 
we are using the energy-independent Thomson optical depth as
length variable, instead of the total optical depth $d\tau(E)=-\chi(z,E)dz$.

The second term in the right-hand side is the source function, which is 
defined at each depth and frequency as
\begin{equation}\label{eqsource}
S(E) = \frac{\eta(E)}{\chi(E)}
= \frac{\alpha_{kn}J(E)+j(E)}{\alpha_{kn} + \alpha_a}
\end{equation}
where
\begin{equation}\label{eqjmom}
J(E) = \int_0^1{u(\mu,E)d\mu},
\end{equation}
is the mean intensity (first moment of the intensity), and $j(E)$ is
the thermal continuum plus line emissivity calculated at each depth.
Here $\alpha_{kn}$ and $\alpha_a$ are the scattering and absorption 
coefficients, respectively: the former is defined by the
product of the density times the Klein-Nishina cross section, while the
latter contains the continuum cross section due bound-bound,
bound-free and free-free absorption. Both $j(E)$ and $\alpha_a$ are obtained 
from the ionization balance at each point using {\sc xstar}.

We also consider the redistribution of the photons due to Compton
scattering using a Gaussian kernel convolved with $J(E)$. The Gaussian is centered at
\begin{equation}\label{eqecen}
E_c = E_0(1+4\theta-\epsilon_0),
\end{equation}
where $\theta=kT/m_ec^2$ is the dimensionless temperature,
$E_0$ is the initial photon energy and $\epsilon_0\equiv E_0/m_ec^2$. The energy dispersion
is given by $\sigma = E_0 \left[ 2\theta + \frac{2}{5}\epsilon_0^2 \right]^{1/2}$.

Therefore the total source function at each depth, as a function of the energy, is given by:
\begin{equation}\label{eqsou}
S(E) = \frac{\alpha_{kn}}{\chi(E)}J_c(E) + \frac{j(E)}{\chi(E)}
\end{equation}
where
\begin{equation}\label{eqjc}
J_c(E) = \int{dE'J(E')P(E',E)}
\end{equation}
is the Comptonized mean intensity, for which the kernel function (normalized to unity), can be written explicitly as
\begin{equation}
P(E_c,E_s) = \frac{1}{\sigma\pi^{1/2}}\exp{\left[ \frac{-(E_s-E_c)^2}{\sigma^2}\right]}.
\end{equation}
As shown by \cite{ros93}, this treatment describes adequately the down scattering
of photons with energies less than $\sim 200$ keV. This approximation is also discussed
by \cite{ros78} and \cite{nay00}. There is an important limitation while using this
treatment: if the bin width of a certain energy $E_c$ is greater than 2 times the dispersion of the Gaussian,
then there is a large probability for a photon to scatter into its own energy bin,
producing a numerical pile up of photons at those energies. We use a logarithmically
spaced energy grid, so the resolving power ${\cal R}=E/\Delta E$ is a constant that only
depends on the number of grid points. Therefore, the bin width is $\Delta E=E/{\cal R}$,
and the pile up occurs at a critical energy $E_p$ when $\Delta E=2\sigma(E_p)$; or
\begin{equation}\label{eqpile}
E_p = m_ec^2 \left[\frac{5}{4{\cal R}^2}-\frac{5kT}{m_ec^2} \right]^{1/2}.
\end{equation}
From the last equation it is clear that the pile up appears only at low temperatures,
specifically for temperatures less than $T_p=m_ec^2/4k{\cal R}^2$. The lowest resolving
power used in this paper is ${\cal R}=350$, and then $T_p\sim 10^{4\ \circ}$K. We will show
later that we consider no cases for which the gas temperature becomes lower than this
value. In fact, in the extreme limit of $T=0$ (and using the same resolving power), 
the highest pile up energy would be $E_p\sim 1.6$ keV, leaving the high energy part of
the spectrum unaffected.

To complete this solution, two boundary conditions are imposed; one at the top ($\tau=0$),
and one at the maximum depth ($\tau=\tau_{max}$). At the top, the 
incoming radiation field $I(0,-\mu,E)$ is equal to a power law, with photon 
index and normalization as free (input) parameters in the calculation.
Subtracting Equations~(\ref{equav}) and (\ref{eqvav})
\begin{equation}\label{eqbc1}
u(0,\mu,E) - v(0,\mu,E) = I_{inc}
\end{equation}
and it is easy to show that
\begin{equation}\label{eqveq}
v(\tau,\mu,E) = \omega(\tau,E) \mu \frac{\partial u(\tau,\mu,E)}{\partial \tau}
\end{equation}
therefore at the surface,
\begin{equation}\label{eqbco}
\omega(0,E) \mu \left[ \frac{\partial u(\tau,\mu,E)}{\partial \tau} \right]_0
- u(0,\mu,E) = - I_{inc}.
\end{equation}
At the lower boundary we specify the outgoing $I(\tau,+\mu,E)$ radiation field to be equal 
to a blackbody with the expected temperature for the disk:
\begin{equation}\label{eqbcm}
\omega(\tau_{max},\mu,E)\mu \left[ \frac{\partial u(\tau,\mu,E)}{\partial \tau} \right]_{\tau_{max}}
+ u(\tau_{max},\mu,E) = B(T_{disk})
\end{equation}
This condition appears since we assume that there is an intrinsic disk radiation due to the 
viscosity of the gas \citep{sak73}. The disk temperature at the lower
boundary is then given by
\begin{equation}\label{eqtd}
T_{disk} = \left\{ \frac{3GM\dot{M}}{8\pi\sigma R^3} \left[ 1 - \left(\frac{R_0}{R}\right)^{1/2} \right] \right\}^{1/4}
\end{equation}
where $G$ is the Newtonian gravitation constant, $\sigma$ is Stefan's constant,
$M$ is the mass of the central object, $\dot{M}$ the mass accretion rate, $R$
is the distance from the center, and $R_0$ is the smallest radius at which the disk dissipates 
energy. \cite{sak73} predicted that for a non-rotating black hole this occurs at 
the innermost stable circular orbit (ISCO), or $R_0=3R_s=6GM/c^2$, where $R_s$ is the Schwarzschild radius. 
However, recent magneto hydrodynamics calculations \citep{nob10}, have shown that the electromagnetic 
stress responsible for the energy generation rises steadily inward in the region inside the ISCO, 
falling sharply to zero just before the event horizon. Therefore, we will assume $R_0=R_s$, 
allowing energy generation in the inner part of the disk. It is clear that the properties of the accretion
disk are introduced in the calculation by means of the lower boundary condition. Further,
since we assume constant density for the gas along the vertical direction, it is customary
to parametrize each model by the ratio of the net flux incident at the surface over the
density, using the common definition of the ionization parameter \citep{tar69}:
\begin{equation}\label{eqxi}
\xi = \frac{4\pi F_{\rm x}}{n_e},
\end{equation}
where $F_{\rm x}$ is the flux of the illuminating radiation in the 1-1000 Ry energy band.
Finally, Equations~(\ref{eqrt}),(\ref{eqsou}),(\ref{eqbco}) and (\ref{eqbcm}) are converted
to a set of difference equations by discretization of depths, energies and angles.
The solution of the system is found by forward elimination and back substitution. A full 
transfer solution requires the Feautrier solution to be iteratively repeated, in order
to self-consistently treat the scattering process. This procedure requires $\sim \tau_{max}^2$
iterations (lambda iterations) for convergence.
%

\subsection{Structure of the gas}\label{secstr}
Given the solution for the radiation field at each point 
in the atmosphere, we use the photoionization code {\sc xstar}
\citep{kal01} to determine the state of the gas for a given value
of the number density.
The state of the gas is defined by its temperature and the
level populations of the ions. The relative abundances of the
ions of a given element and the level populations are found by solving the ionization
equilibrium equations under the assumption of local balance,
subject to the constrain of particle number conservation for
each element. Schematically, for each level
\begin{equation}\label{eqion}
\mathrm{Rate\ in = Rate\ out}
\end{equation}
The processes include spontaneous decay, photoionization, charge transfer,
electron collisions, radiative and dielectronic recombination.
Similarly, the temperature of the gas is found by solving the equation of
thermal equilibrium, which may be written schematically as
\begin{equation}\label{eqhea}
\mathrm{Heating = Cooling}
\end{equation}
The heating term includes photoionization heating (including
the Auger effect), Compton heating, charge transfer and collisional
de-excitation. The cooling term includes radiative and dielectronic
recombination, bremsstrahlung, collisional ionization, collisional
excitation of bound levels and charge transfer. All processes include
their respective inverses so that the populations approach to local
thermodynamic equilibrium (LTE) values under the appropriate conditions
(i.e., high density or Planckian radiation field).
%

\subsection{Radiative equilibrium}\label{secreq}
{\sc xstar} calculates level populations, temperature, opacity and 
emissivity of the gas assuming that all the physical processes mentioned
in the previous section are in steady state. Radiative equilibrium is
achieved by calculating the integral over the net emitted and absorbed
energies in the radiation field ($E_c$ and $E_h$, respectively); and 
varying the gas temperature until the integrals satisfy the criterion
\begin{equation}\label{eqhc}
\frac{E_h - E_c}{E_h + E_c} \leq 10^{-4}.
\end{equation}
The radiative equilibrium condition must be ensured also while 
solving the transfer equation. According to equation~(2-83b) in
\cite{mih78}:
\begin{equation}\label{eqradeq}
\int_0^{\infty}{\chi(E)\left[J(E) - S(E) \right] dE} = 0,
\end{equation}
i.e., the total energy absorbed by a volume of material must be equal
to the total energy emitted (note that, as in \S~\ref{secrad}, we have
omitted the explicit dependence on $\tau$ in the equations).
By using the definitions of the flux and radiation pressure (second
and third moments of the radiation field):
\begin{equation}\label{eqhmom}
H(E)=\int_0^1{v(\mu,E)\mu d\mu} 
\end{equation}
and
\begin{equation}
K(E)=\int_0^1{u(\mu,E)\mu^2 d\mu},
\end{equation}
one can rewrite Equation~(\ref{eqveq}) as
\begin{equation}
H(E) = \omega(E)\frac{\partial K(E)}{\partial\tau}.
\end{equation}
Taking the derivative with respect to $\tau$ and multiplying
by $\omega(E)$
\begin{equation}
\omega(E)\frac{\partial H(E)}{\partial\tau} = \omega^2(E)\frac{\partial^2K(E)}{\partial\tau^2}
+ \omega(E)\frac{\partial\omega(E)}{\partial\tau}\frac{\partial K(E)}{\partial\tau},
\end{equation}
and comparing with the transfer Equation~(\ref{eqrt}) integrated over $\mu$,
\begin{equation}\label{eqreq1}
\omega(E)\frac{\partial H(E)}{\partial\tau} = J(E) - S(E).
\end{equation}
From here it is clear that the radiative equilibrium condition~(\ref{eqradeq})
is equivalent to
\begin{equation}\label{eqdh}
\frac{\partial}{\partial\tau}\int_0^{\infty}{H(E) dE} = 0,
\end{equation}
i.e., the net flux must be conserved for any optical depth. Further, 
inserting the source function~(\ref{eqsou}) in the right-hand side of~(\ref{eqreq1}):
\begin{equation}\label{radeq2}
\frac{\partial}{\partial\tau}\int_0^{\infty}{H(E) dE} = 
\frac{1}{\alpha_T}\int_0^{\infty}{\left\{ \alpha_{kn}\left[J(E)-J_c(E) \right] + \alpha_aJ(E) - j(E) \right\} dE}.
\end{equation}
If the gas temperature is high enough ($T \gtrsim 10^{6\ \circ}$K), the Compton 
scattering is the dominant process, while both the opacity and the emissivity of
the gas are negligible and only the first term in the right-hand side of 
Equation~(\ref{radeq2}) is important. For lower temperatures the inverse occurs:
$J(E)\approx J_c(E)$, which cancels out the first term. Thus, using~(\ref{eqdh})
\begin{equation}\label{radeq3}
\frac{1}{\alpha_T}\int_0^{\infty}{\left[\alpha_aJ(E) - j(E) \right] dE}=0.
\end{equation}
The conservation of the flux through the atmosphere is formally equivalent to
the thermal equilibrium condition (\ref{eqhea}), but its accuracy will depend on
the error associated with the evaluation of the left-hand side of 
Equation~(\ref{radeq3}). Although we require a small error in the calculation of
the emissivities $j(E)$ and opacities $\alpha_a$ by imposing the condition~(\ref{eqhc}),
the numerical conservation of the flux is a difficult task due to the fact that
small errors accumulate over the large column densities covered in the calculation.
It can be seen from Equation~(\ref{radeq3}) that the error required in the calculation
of emissivities and opacities must be of the order of $\alpha_T^{-1}$. Assuming that 
typical values for the density are $n_e\sim 10^{10} - 10^{16}$ cm$^{-3}$,
then $10^8 \lesssim \alpha_T^{-1} \lesssim 10^{14}$, much greater than the
value required in (\ref{eqhc}). Therefore, we also apply a renormalization
of the emissivities calculated by {\sc xstar} before the solution of the
transfer equation, such that:
\begin{equation}
j^{new}(E) = j(E) \frac{\int_0^{\infty}{\alpha_a J(E)dE}}{\int_0^{\infty}{j(E)dE}},
\end{equation}
at each depth. By implementing this correction in our models, we have found 
that the net flux (and therefore the energy) is conserved better than 1\% for intermediate
to large values of the ionization parameter (log~$\xi > 3$), but the maximum error can be 
$\sim 7\%$ for low-ionization calculations (log~$\xi<3$). These errors are estimated by
comparing the total fluxes incident on the gas (i.e., $F_{\mathrm x} + F_{disk}$), with
the total emergent flux.
%

\subsection{Atomic data}\label{secato}
The {\sc xstar} atomic database collects recent data from many sources 
including CHIANTI \citep{lan06}, ADAS \citep{sum04}, NIST \citep{ral08}, 
TOPbase \citep{cun93} and the IRON project \citep{hum93}. The database
is described in detail by \cite{bau01}. Additionally, the atomic data 
associated with the K-shell of the Fe ions incorporated 
in the current version of {\sc xstar} has been recently calculated and
represents the most accurate and complete set available to the present.
Energy levels and transition probabilities for first row ions 
Fe~{\sc xviii}-Fe~{\sc xxiii} were reported by \cite{pal03a}; for 
second row ions Fe~{\sc x}-Fe~{\sc xvii} were reported by \cite{men04};
and for the third row ions Fe~{\sc ii}-Fe~{\sc ix} by \cite{pal03b}.
The impact of the damping by spectator Auger resonances on the 
photoionization cross sections was discussed by \cite{pal02}. Photoionization
and electron impact cross sections were presented for second
row ions by \cite{bau04}. Energy levels, transition probabilities and
photoionization cross sections for Fe~{\sc xxiv} were calculated by
\cite{bau03}. A compilation of these results and a careful study of
their impact on the photoionization models can be found in \cite{kal04}.
Moreover, {\sc xstar} also includes the atomic data relevant to the 
photoabsorption near the K edge of all oxygen ions calculated by
\cite{gar05}.

The approach used by these authors for the computation of the atomic parameters
of iron and oxygen was based on the implementation and comparison of results
obtained from different atomic codes of public domain, namely 
{\sc autostructure} \citep{bad97}, {\sc hfr} \citep{cow81}, and the 
Breit-Pauli {\it R}-Matrix package \citep[{\sc bprm},][]{sea87, ber87}.
%

\subsection{Iteration procedure}\label{secite}
Starting at the top of the disk, the vertical structure 
of the gas is found by solving ionization and thermal balance at 
each spatial zone using {\sc xstar}, as described in \S~\ref{secstr}. 
Since the radiation field is unknown the first time this is done, 
it is set to be equal to the incident power law at each depth. 
Once the last zone is reached, the temperature and 
density profiles are known, as well as the emissivities and opacities 
for each depth and energy. With this information the radiative transfer 
Equation~(\ref{eqrt}) is solved as described in \S~\ref{secrad} until the 
solution converges, which requires $\sim \tau_{max}^2$ Lambda iterations. 
This provides a new and more accurate radiation field, that can be used 
to recalculate the structure of the gas. Because the emissivities and
opacities are updated when the structure of the gas is recalculated,
the radiative transfer calculations must be also repeated. We then continue
with this process until all quantities stop changing within a 
small fraction. Typically, this requires $\sim 20$ gas structure 
calculations, times $\tau_{max}^2$ Lambda iterations.

For our purposes, we desire to cover the largest optical depth possible
for realistic computational times and resources. Since Lambda iteration requires
$\tau_{max}^2$ loops to converge, we performed different numerical experimentation
changing the value of $\tau_{max}$. Comparisons with the semi-analytic solution 
by \cite{cha60} showed that the error behaves, in general, as $\sim 1/\tau_{max}$,
corresponding to a leakage of energy through a scattering dominated slab.
The comparison also showed that diminishing improvement is achieved after $\tau_{max} \sim 10-20$,
taking into account the required number of iterations. Therefore we limit all our
calculations to $\tau_{max}=10$, with the understanding that our resulting global energy
budget may be in error by as much as 10\%. Finally, our calculations consider high resolution
spectra with an energy grid of at least $5\times10^3$ points (${\cal R}=E/\Delta E\sim 350$),
200 spatial zones, and 50 angles to account for anisotropy of the radiation field.
Typically, each one of these models can be calculated in few hours in a last
generation PC, which means that a grid of 20-30 models can be produced in one
or two weeks, depending on the computational resources available, and the specific
parameters to represent the models (e.g., lower ionization models tend to run slower
than high ionization cases). This is relevant since {\sc xspec} model tables 
can be easily generated and used in
the interpretation of astronomical observations. Higher resolution can be 
achieved, although computational time increases (at least) with the square 
of the number of energy grid points, quickly imposing limits on the sizes
of the grids.

%
%
\section{Results}\label{secres}
In this section we show the results obtained with our model for constant density atmospheres.
We present a total of 20 reflection models, calculated for various representative conditions. These models
and input parameters used to produce them are summarized in Table~\ref{tamodels}.
The first column of the Table contains an identification number assigned to each model 
and it will be used for reference through the rest of this paper. The next columns contain the
input parameters that have been varied for each model, namely: the flux $F_{\rm x}$ of the ionizing radiation
(in the 1-1000 Ry energy range), the resulting logarithm of the ionization parameter $\xi$ (see Equation~\ref{eqxi}),
the spectral energy resolution or resolving power ${\cal R}=E/\Delta E$ (determined by the number of energy
grid-points used), the cosine of the incidence angle of the illuminating radiation with respect to
the normal of the disk $\mu_0$, and the abundance of iron normalized to its
solar value A$_{\rm Fe}$. Other input parameters not listed in the Table since they
are constant and common to all the models are: density $n_e=10^{15}$ cm$^{-3}$, 
photon index of the incident radiation $\Gamma=2$, mass of the central object $M=10^8$ M$_{\sun}$, distance
from the central object $R=7 R_s$, and the mass accretion rate 
$\dot{M}=1.6\times 10^{-3} \dot{M}_{Edd}$, where $R_s=2GM/c^2$ is the Schwarzschild radius and
$\dot{M}_{Edd}$ is the accretion rate at the Eddington limit. We have chosen these
parameters such that the numbers are similar to those typically used by previous
authors. However, note that these parameters only affect the present models by changing
the effective temperature of the disk (equivalent to changing the intrinsic disk flux), to
be used as the inner boundary condition (\ref{eqbcm}) in the radiation transfer problem.
Specifically, by using Equation~(\ref{eqtd}) this set of parameters
correspond to an intrinsic disk flux of $F_{disk}=3.6\times 10^{13}$ erg/cm$^2$/s, or 
an effective temperature of $T_{disk}=2.8\times 10^{4\ \circ}$K, which represents a
cold disk when compared to the temperatures typically produced by the illuminating fluxes used in this
paper. This is convenient since it allows us to analyze the reprocessed spectrum as a
direct consequence of the incident power law without significant modifications due to the
black body of the disk.
\begin{deluxetable}{cccccc}
\tabletypesize{\scriptsize}
\tablecaption{List of reflection models with their respective input parameters
\label{tamodels}}
\tablewidth{0pt}
\tablehead{
\colhead{Model} & \colhead{$F_{\rm x}$\tablenotemark{a}} & \colhead{log~$\xi$\tablenotemark{b}} & \colhead{${\cal R}$} & \colhead{$\mu_0$} & \colhead{A$_{\rm Fe}$}\\
}
\startdata
1  &  $5\times 10^{14}$     &  0.8  & 350  &  0.71  &  1.0 \\
2  &  $1\times 10^{15}$     &  1.1  & 350  &  0.71  &  1.0 \\
3  &  $2\times 10^{15}$     &  1.5  & 350  &  0.71  &  1.0 \\
4  &  $5\times 10^{15}$     &  1.8  & 350  &  0.71  &  1.0 \\
5  &  $1\times 10^{16}$     &  2.1  & 350  &  0.71  &  1.0 \\
6  &  $2\times 10^{16}$     &  2.5  & 350  &  0.71  &  1.0 \\
7  &  $5\times 10^{16}$     &  2.8  & 350  &  0.71  &  1.0 \\
8  &  $1\times 10^{17}$     &  3.1  & 350  &  0.71  &  1.0 \\
9  &  $2\times 10^{17}$     &  3.5  & 350  &  0.71  &  1.0 \\
10  &  $5\times 10^{17}$    &  3.8  & 350  &  0.71  &  1.0 \\
11  &  $2.5\times 10^{15}$  &  1.5  & 3500 &  0.71  &  1.0 \\
12  &  $2.5\times 10^{16}$  &  2.5  & 3500 &  0.71  &  1.0 \\
13  &  $2.5\times 10^{17}$  &  3.5  & 3500 &  0.71  &  1.0 \\
14  &  $5\times 10^{16}$    &  2.8  & 350  &  0.95  &  1.0 \\
15  &  $5\times 10^{16}$    &  2.8  & 350  &  0.50  &  1.0 \\
16  &  $5\times 10^{16}$    &  2.8  & 350  &  0.05  &  1.0 \\
17  &  $2.5\times 10^{16}$  &  2.5  & 350  &  0.71  &  0.2 \\
18  &  $2.5\times 10^{16}$  &  2.5  & 350  &  0.71  &  2.0 \\
19  &  $2.5\times 10^{16}$  &  2.5  & 350  &  0.71  &  5.0 \\
20  &  $2.5\times 10^{16}$  &  2.5  & 350  &  0.71  & 10.0 \\
\enddata
\tablenotetext{a}{erg/cm$^{2}$/s (1-1000 Ry)}
\tablenotetext{b}{erg cm/s (See Equation~\ref{eqxi})}
\end{deluxetable}
%
%
\subsection{Temperature profiles}\label{sectem}
\begin{figure}
\epsscale{1.}\plotone{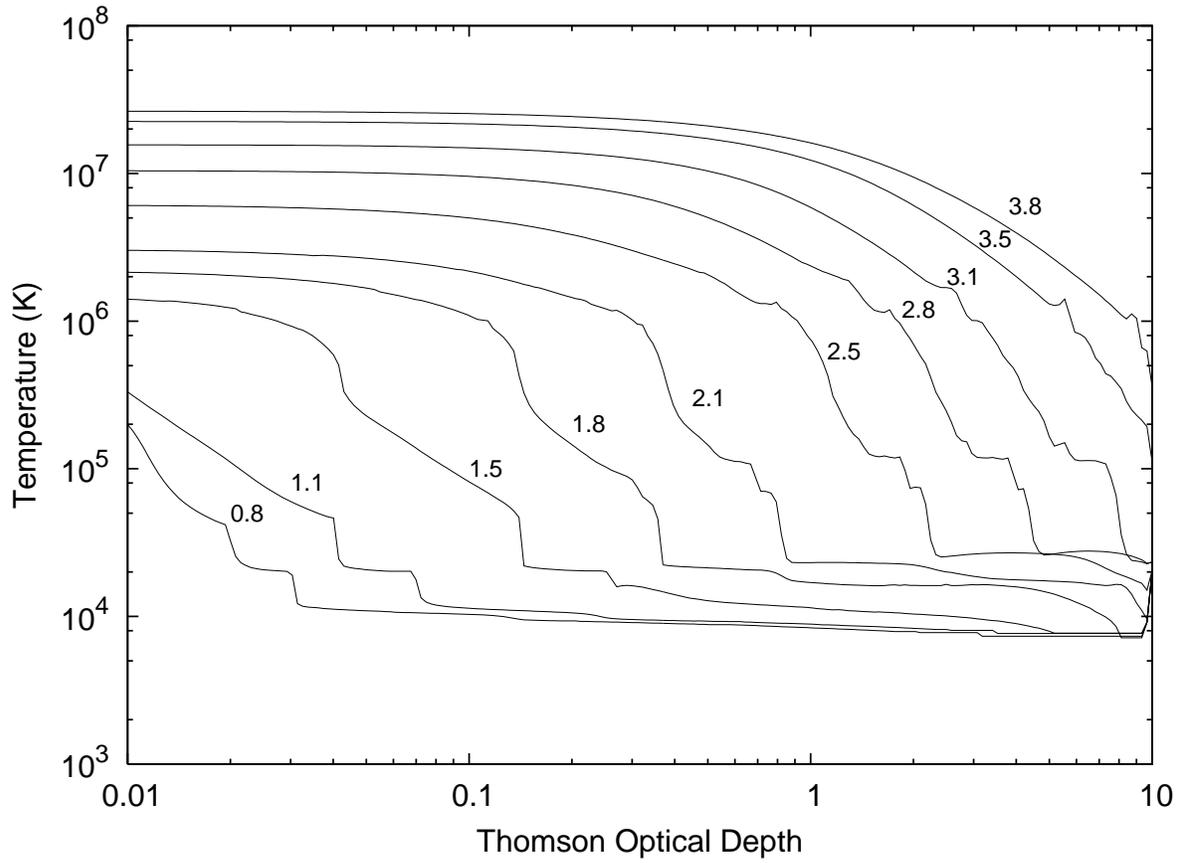}
\caption{Temperature profiles for different illumination fluxes using
constant density models ($n=10^{15}$ cm$^{-3}$). The value of log~$\xi$ 
is shown next to each corresponding curve (models 1-10).\label{temp}}
\end{figure}

The Figure~\ref{temp} shows the temperature profile as a function of the
Thomson optical depth resulting from constant density models for ten
different ionization parameters (models 1-10 in Table~\ref{tamodels}). 
In the Figure, the lower left curve corresponds to the least ionized case, 
and each consecutive to a higher value of $F_{\rm x}$. The values of log~$\xi$
are included next to the respective curves. Because the 
$\tau$ grid is fixed with logarithmic spacing, the surface of the disk is in 
practice chosen to be $\tau_{top}=10^{-2}$. The optical depth is measured
from the surface towards the interior of the disk. Since the density is
assumed to be constant, all these calculations are carried out along a
distance of $\Delta z = \Delta\tau/\sigma_T n_e \approx 1.5\times 10^{10}$ cm.
 
These curves show very similar general behavior; the temperature
is higher at the surface (due to the heating by the incident radiation)
and decreases towards the interior of the disk, reaching a minimum
temperature. For the high illumination cases (large values of $\xi$, 
and/or very close to the surface), the dominant mechanism is Compton
heating and cooling \citep{kro81}, with a rate given by:
\begin{equation}\label{eqheat}
n_e\Gamma_e = \frac{\sigma_T}{m_ec^2}\left[\int{EF(E)dE} - 4kT\int{F(E)dE}\right],
\end{equation}
in the non relativistic limit. When this process dominates, the temperature approaches an asymptotic 
value, the {\it Compton temperature}, given by the balance of the two terms in the previous equation:
\begin{equation}\label{eqtic}
T_{\rm C}=\frac{<E>}{4k},
\end{equation}
where 
\begin{equation}\label{eqmean}
<E> = \frac{\int{F(E)EdE}}{\int{F(E)dE}}=F_{\rm tot}^{-1}\int{F(E)EdE}
\end{equation}
is the mean photon energy. From Equations~(\ref{eqtic}) and (\ref{eqmean}) it
is clear that $T_{\rm C}$ only depends on the shape of the spectrum. In particular,
for the incident power law used in this paper ($\Gamma=2$), the Compton
temperature of the radiation is $T_{\rm C}=2.8\times 10^{7\ \circ}$K. Note that
the temperature at surface for the case with log~$\xi=3.8$ (model 10) is indeed
very close to $T_{\rm C}$.

In most cases, the gas remains at the high temperature limit throughout the region where the
ionizing radiation field is unchanged from the incident field, thus forming
a hot skin. The temperature gradually decreases in this hot skin as the 
bremstrahlung cooling becomes more important. As photons are removed from the
Lyman continuum energies either by scattering or absorption and re-emission, 
at some point hydrogen recombines and the opacity grows rapidly, causing
a drop in the temperature. Eventually the opacity becomes very large
so that few photons are left in the 1-1000 Ry energy range, at which point the
radiation field thermalizes to a nearly constant temperature ($T \sim 10^{4\ \circ}$K).

It is worthwhile to mention that for the models with the two lowest ionization
parameters (models 1 and 2), the opacity of the gas is large compared to 
the ionizing flux even at the surface of the disk and therefore the gas reaches
the low temperature limit very rapidly. Conversely, for the two models
with log~$\xi=2.5-3.1$ (models 6-8), the illumination is high enough to allow
the photons to penetrate much deeper into the atmosphere, and the gas remains
at high temperature ($T\sim 10^{6\ \circ}$K) even for large optical depths
($\tau > 1$). In fact, by looking the intermediate cases (models 3-8),
one can clearly identify two temperature zones: the hot skin correspond
to $T \gtrsim 10^{6\ \circ}$K, while the cold region occurs for 
$T \lesssim 10^{5\ \circ}$K. Note also that although the transition between the
hot and cold regions can be sudden, there is no thermal instability in these
calculations, since in all these cases the constant density restriction is applied
\citep{kro81}. Some of the models show small temperature inversions, such as 
the one in the curve correspoding to log~$\xi=3.5$ (model 9), that can be seen 
around $\tau_T \approx 6$. This inversion occurs in the place within the slab 
where the temperature changes 
rapidly but also in a zone where the resolution in the optical depth is reduced 
due to larger the step size in the grid, which could lead to small scale thermal
instabilities in the solution \citep{buf74}. Despite this, there are no large
scale thermal instabilities in these calculations, since in all these cases the 
constant density restriction is applied \citep{kro81}. 
%
%
\subsection{Reflected spectra}\label{secref}
\begin{figure*}
\epsscale{1.0}\plotone{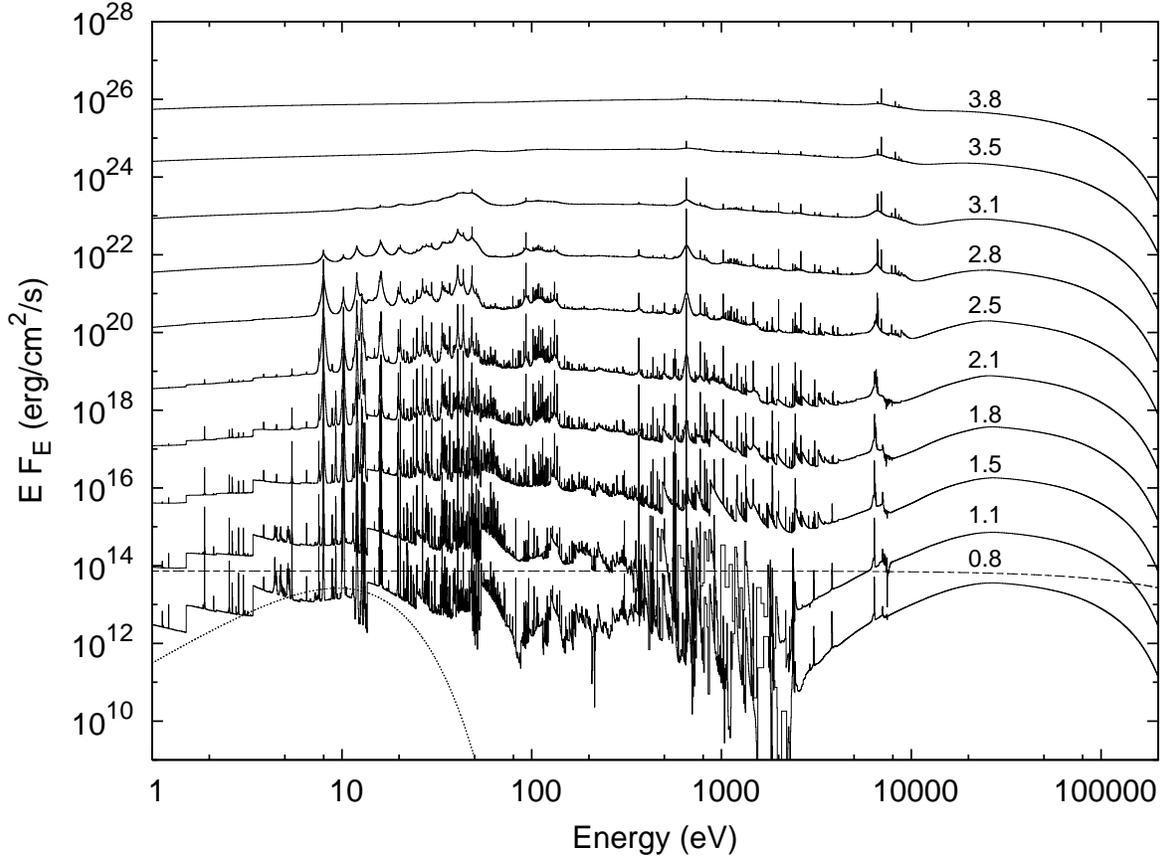}
\caption{{\it Solid curves:} Reflected spectra for the models presented in Figure~\ref{temp}.
The value of log~$\xi$ is shown next to each corresponding curve
(models 1-10). The curves are shifted by arbitrary factors for clarity.
These are, from bottom to top: 1, 10, 100,...,$10^{9}$. {\it Dashed curve:} incident ionizing
spectrum (power law), corresponding to the lowest ionization case (log~$\xi=0.8$). {\it Dotted curve:}
intrinsic disk flux (black body), common to all the models.}
\label{spec}
\end{figure*}

Figure~\ref{spec} shows the reflected spectra for each of the models shown
in the previous section (models 1-10 in Table~\ref{tamodels}), in the entire 
energy range covered in the calculations (1 eV to 210 keV). 
The spectra emerging from the top of the slab are plotted
as solid curves, while the intrinsic disk flux is shown in the dotted curve. 
The dashed curve represents the X-rays incident at the top of the disk,
assumed to be in the form of a power law with $\Gamma=2$ and an exponential high
energy cutoff at 200 keV (to improve clarity, only the incident power law for
the lowest ionization case is shown).
The corresponding values of log~$\xi$ are indicated above each curve, starting
with the smaller value at the bottom, and increasing the ionization to the top. The 
curves are shifted to improve clarity. From bottom to top,
each consecutive curve is rescaled by a factor increased by one order of magnitude
with respect to the previous one (i.e., 1, 10, 100,...,$10^{9}$). 
Strong absorption profiles and edges are clearly seen in the neutral cases for
log~$\xi=$0.8, 1.1 and 1.5 (models 1-3), although it is 
important to mention that some small numerical problems occur in the thermal
energy range of the spectrum for the lower flux case. For a few energy bins the 
reflected outgoing flux becomes negative, due to the large opacities
for such energies (especially around 100 eV) and numerical errors in calculating
the mean intensity. The redistribution of the photons due to Comptonization is evident above 10 keV (Compton bump),
and in the smearing of the line profiles. For the models with the lowest illumination
(models 1-3), there is a significant modification of the original power law 
continuum due to the large values of the photoelectric opacity for energies between
100~eV and 10~keV, where most of the strong absorption occurs. Nevertheless, none
of these models shows a spectrum dominated by absorption. Even in the lowest 
ionization case (log~$\xi=0.8$), there are strong emission lines present through
the whole energy range. This combination of emission and absorption features in
the reflected spectra is a direct consequence of our accurate treatment of
radiation transfer and the temperature gradient along the
vertical direction of the disk discussed in the previous section. 

The iron K line shows the effects of Compton scattering in models 
with log~$\xi \ge 1.5$, since those are the cases for which a hot skin ($T\sim10^{6\ \circ}$K)
is present for at least a fraction of the total depth of the disk.
By comparing the most ionized 
models, in particular those for log~$\xi=2.8$ and log~$\xi=3.1$ (models 7 and 8), 
one can see a very drastic change from a highly ionized to an almost featureless 
spectrum, and even fewer features are seen in the two models for log~$\xi=3.5$ and
3.8 (models 9 and 10). This is due to the fact that in these cases the gas is 
always at high temperature within the range of our calculations. Despite this, 
emission from highly ionized iron K lines is still apparent in the reflected spectrum.
 
\begin{figure*}
\epsscale{1.0}\plotone{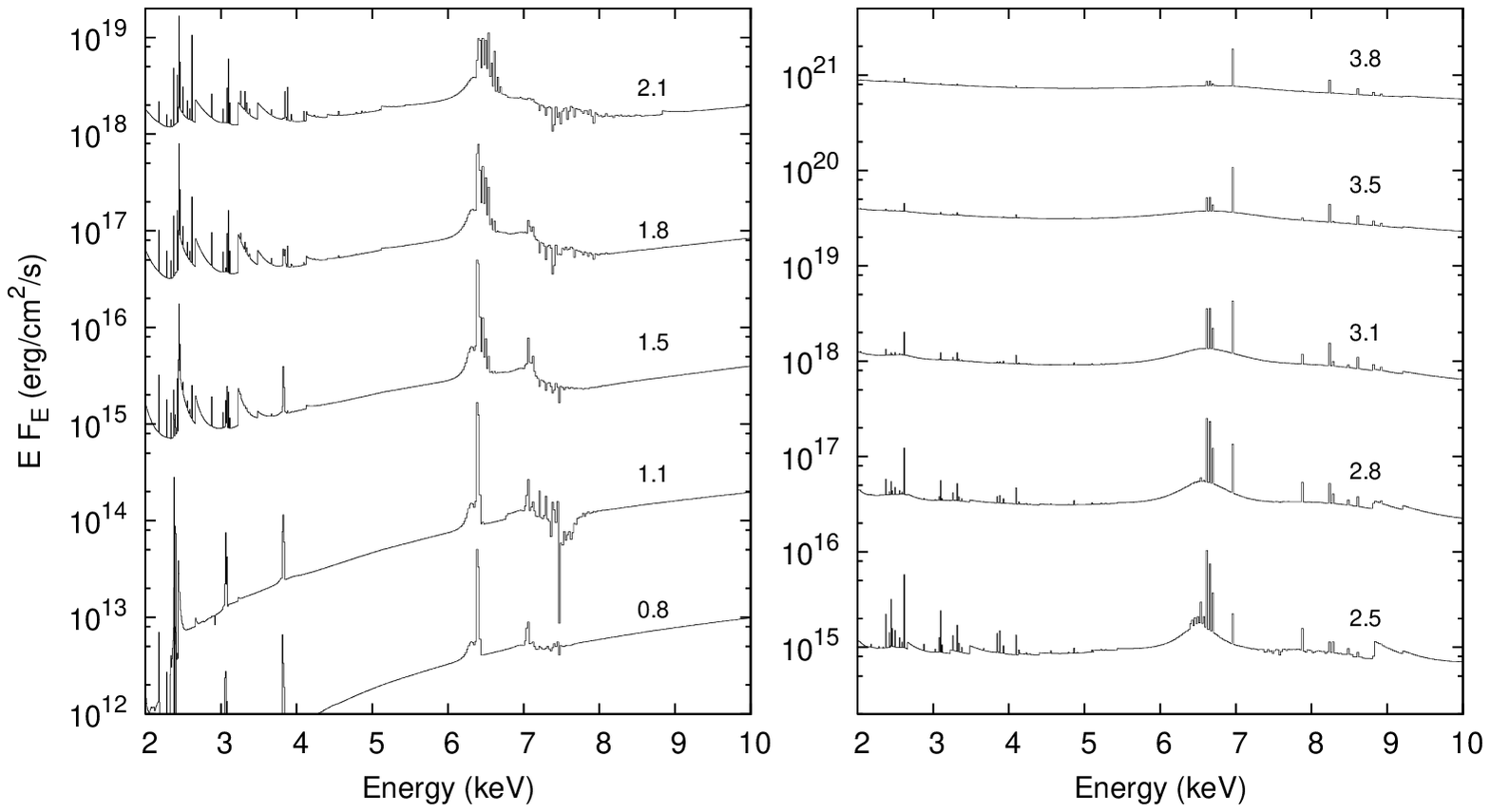}
\caption{Reflected spectra as shown in Figure~\ref{spec} in the 2-10 keV
region. The value of log~$\xi$ is shown next to each
corresponding curve (models 1-10). The curves are shifted by arbitrary
factors for clarity.  These are, from bottom to top in each panel:
1, 10, 100, $10^{4}, 10^5$.}
\label{zoom}
\end{figure*}

Figure~\ref{zoom} shows the same results as Figure~\ref{spec}, but in
the 2-10 keV energy range. Here it is possible to see the structure of the
iron inner shell transitions in detail. The Compton shoulder is apparent
in the first three curves from bottom to top (least ionized cases). Changes
in the Fe K$\alpha$ emission line are distinguishable as the ionization parameter
increases. The neutral K$\beta$ line is only evident in the three least ionized
cases (models 1-3), as well as the Fe absorption edge at $\sim 7.5$ keV.
This figure also shows a more complex structure of the iron line in intermediate
cases of ionization: line positions and intensities vary between mostly 
neutral emission at 6.4 and 7.2 keV for K$\alpha$ and K$\beta$, to a very ionized 
profile with lines around 6.5-7.0 keV and 7.8-8.2 keV for the same transitions.

We have calculated the equivalent widths for the Fe K$\alpha$ lines in all
these spectra. Because of the large deviations of the reprocessed continuum from 
the incident power law in some of these models (particularly for low $\xi$), in order
to calculate the equivalent width we define a local continuum by interpolating a
straight line between two points in the region around the line, specifically
between 6 and 7 keV. The integration is performed within this range as well, 
such that only K$\alpha$ emission is taking into account.

The resulting equivalent widths for the Fe K$\alpha$ emission line vary between 
$\sim 400-800$ eV for the cases with $1 \lesssim \mathrm{log}\xi \lesssim 3$
(left panel in Figure~\ref{zoom}). For higher values of the ionization parameter
the equivalent widths decrease very rapidly, to approximately 40 eV for the most
ionized case (log~$\xi=3.8$, model 10). Since the gas is more ionized for high
illumination, the emission of the line becomes exclusively due to H- and 
He-like iron ions, which combined with the extreme Compton scattering of the 
photons results in the reduction of the equivalent width of the line. This 
tendency resembles the X-ray Baldwin effect \citep{iwa93}, which have 
been observed in many active galactic nuclei spectra \citep[e.g.,][]{pag04}.  
%
%
\subsection{Spectral features}\label{secspe}
\begin{figure*}
\epsscale{1.0}\plotone{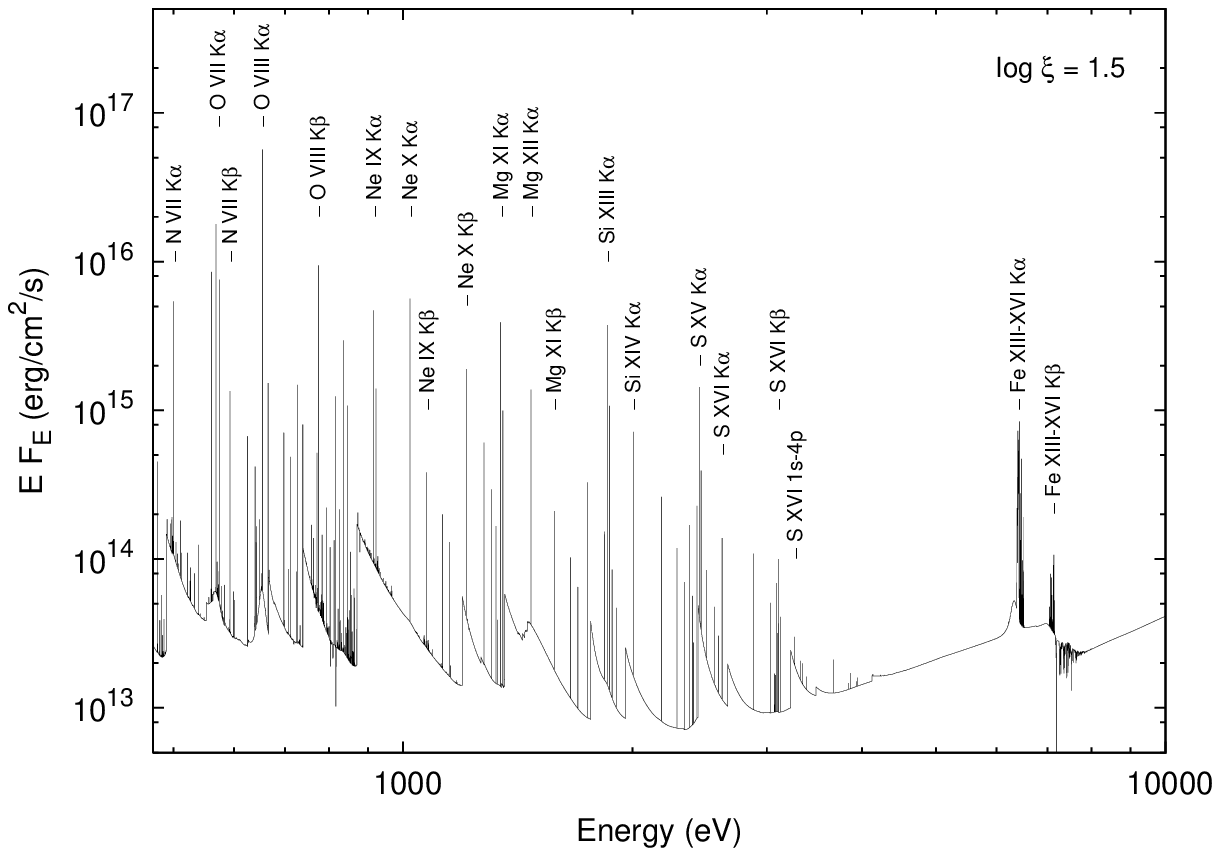}
\caption{Reflected spectra for log~$\xi=1.5$ while using a resolving power of ${\cal R}\sim 3500$ 
(model 11). All the other input parameters are the same used before
(as in Figures~\ref{temp} and \ref{spec}), although no rescaling is applied.
The strongest emission lines are labeled.}
\label{sp18}
\end{figure*}

\begin{figure*}
\epsscale{1.0}\plotone{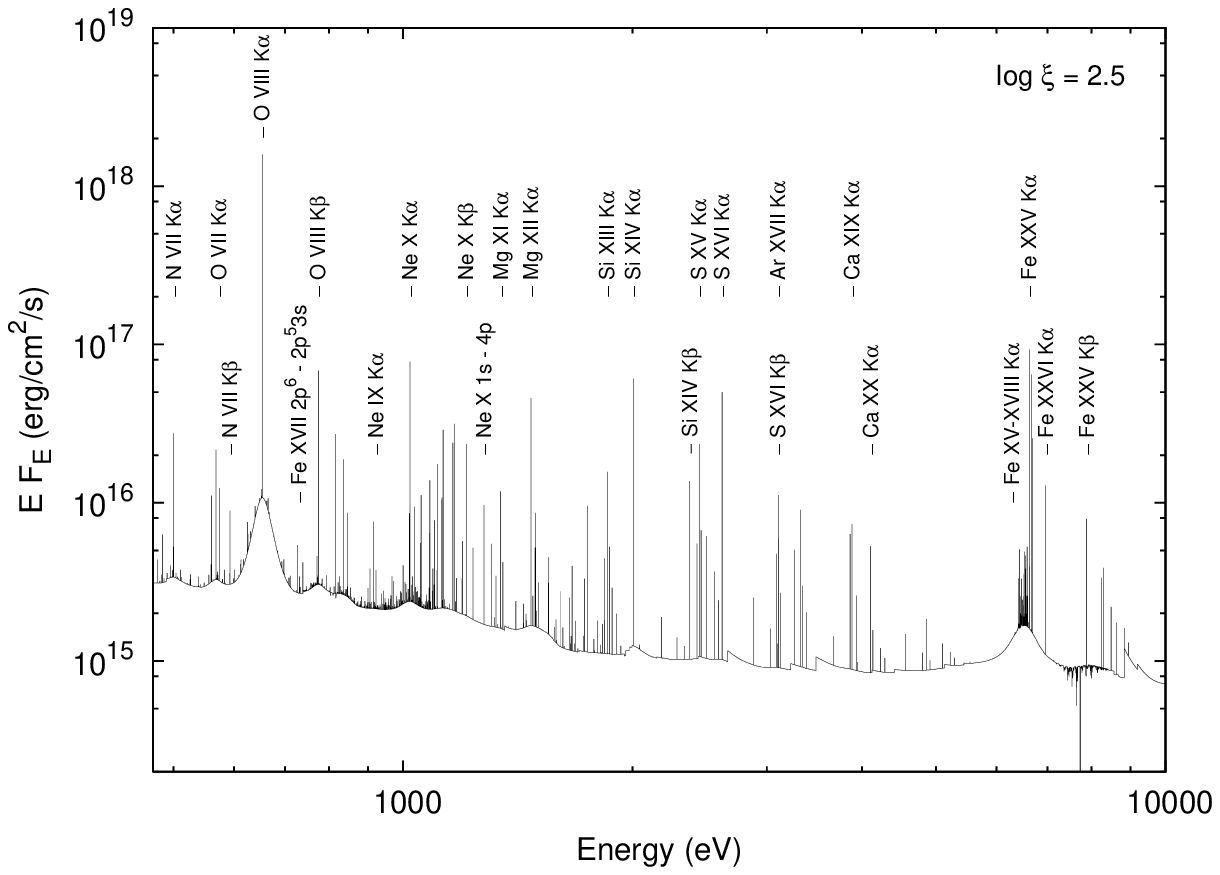}
\caption{Reflected spectra for log~$\xi=2.5$ while using a resolving power of ${\cal R}\sim 3500$ 
(model 12). All the other input parameters are the same used before
(as in Figures~\ref{temp} and \ref{spec}), although no rescaling is applied.
The strongest emission lines are labeled.}
\label{sp28}
\end{figure*}

\begin{figure*}
\epsscale{1.0}\plotone{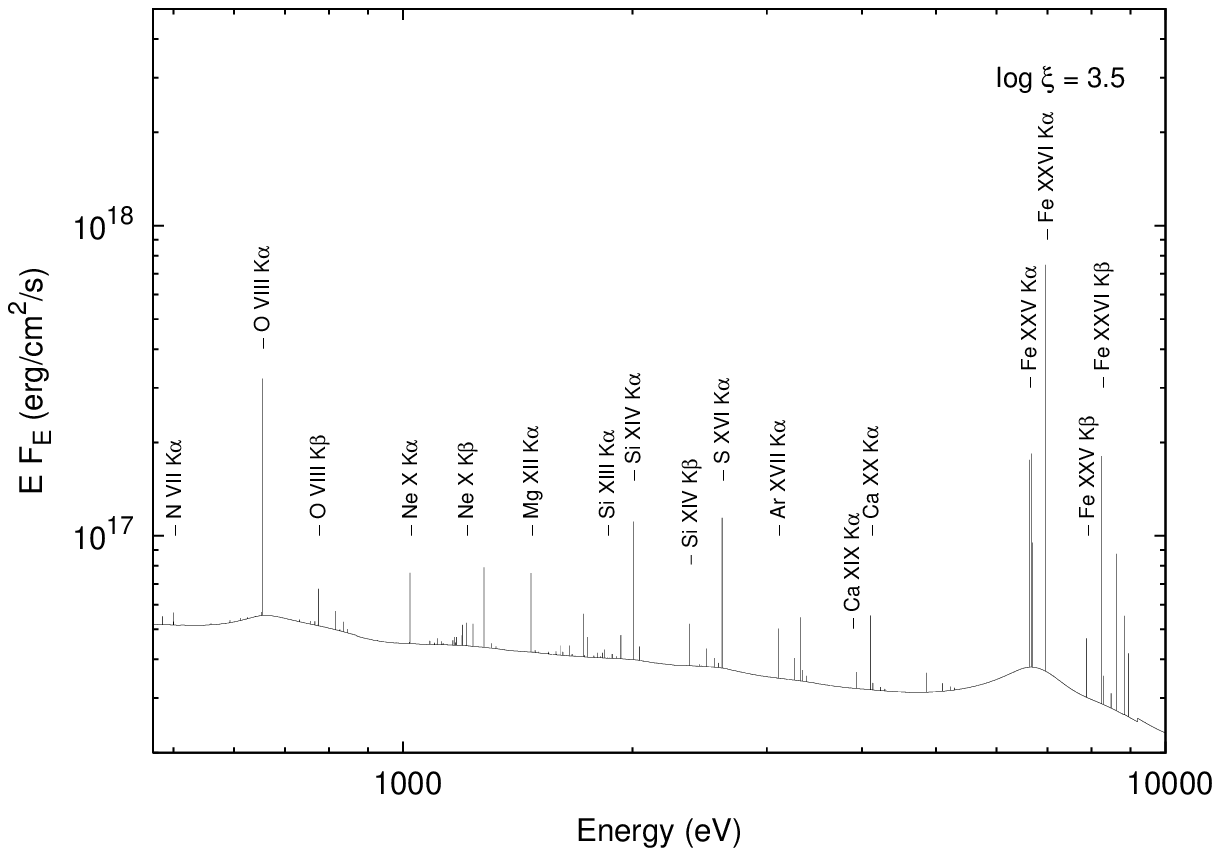}
\caption{Reflected spectra for log~$\xi=3.5$ while using a resolving power of ${\cal R}\sim 3500$
(model 13). All the other input parameters are the same used before
(as in Figures~\ref{temp} and \ref{spec}), although no rescaling is applied.
The strongest emission lines are labeled.}
\label{sp38}
\end{figure*}

In order to provide a finer analysis of the reflected spectra, we perform 
calculations increasing the energy resolution by an order of magnitude, i.e., using
$5\times 10^4$ energy grid-points (${\cal R}\sim 3500$). Since these calculations are more expensive in terms
of CPU-time, we only show three representative cases. These calculations are 
shown in Figures~\ref{sp18}-\ref{sp38}, and correspond to models 11, 12 and 13 in
Table~\ref{tamodels}. These spectra (Figures~\ref{sp18}-\ref{sp38}), are plotted 
in their physical units and without any renormalization or rescaling, and
besides the much higher energy resolution, all the other input parameters in the models are
the same as the ones used for the models present previously in Figures~\ref{temp} and \ref{spec}.

Figure~\ref{sp18} shows the reflected spectra corresponding to log~$\xi=1.5$ (model 11), 
in the 0.5-10 keV energy range, while using a resolving power of ${\cal R}\sim3500$.
The most prominent and relevant emission lines are identified in the Figure. 
In general, the spectrum is dominated by many absorption edges plus radiative recombination 
continua (RRC). RRCs occurs when a free electron is captured by an ion into an
unoccupied orbit; if the electron carries more energy than it needs to be bound to
the ion, the excess will be radiated as a photon. 
Emission K lines from H- and He-like nitrogen, oxygen, neon and magnesium 
are present in the region around and below 2 keV.  At higher energies,
there is clear emission from Si~{\sc xii-xxiv} and S~{\sc xv-xvi}.
The iron K${\alpha}$ and K${\beta}$ components are clearly visible at $\sim 6.4$ keV and
$\sim 7.1$ keV, respectively, as expected from mostly neutral emission. Many lines are blend
together in a small region (about 50 eV for each component), but the emission is mainly 
due to Fe~{\sc xiii} up to Fe~{\sc xvi}. The Compton shoulder in both components is also evident,
as well as a very marked absorption edge at $\sim 7.5$ keV blended with a complex structure
of absorption profiles. These absorption features correspond to resonances given
by transitions from the ground state to the $1s-np$ autoionizing states, where $n=3,...,30$;
for our particular atomic data set. These are K-vacancy states that can decay either 
radiatively or via Auger spectactor channels, being the latter dominant for 
$n \geq 3$ \citep{pal02}. The resonance structure seen in this spectrum covers the 7.1-8. keV 
energy range, thus it is mainly due to second row Fe ions,
which resemble the opacity curves shown by \cite{pal02}, as well as the photionized spectra in \cite{kal04}, 
for similar values of the ionization parameter considered here. The energy position of the absorption 
edge indicates that is likely to be produced by Fe~{\sc xiii-xiv}, according to the energies 
presented in \cite{kal04} (see in particular their Figure~2).

Figure~\ref{sp28} shows the resulting spectra when log~$\xi=2.5$ (model 12), again for a high resolution
energy grid (${\cal R}\sim 3500$). There is a significant change in the overall continuum, particularly
in the 1-10 keV region, in comparison with the previous case (log~$\xi=1.5$). This is
because in the lower ionization case the photoelectric opacity dominates in this energy band
and changes the original incident power law shape, while the continuum spectrum in Figure~\ref{sp28}
still shows the original slope ($\Gamma=2$). Lower opacity also allows the efficient propagation
of lines (since photons are able to escape), and the emission from heavier elements (Ar~{\sc xvii}, Ca~{\sc xix} 
and Ca~{\sc xx}), since the gas is more ionized. In general, this spectrum shows 
more emission lines and weaker absorption
edges. However, some RRCs can still be identified, especially around the sulfur and calcium lines.
The K lines from nitrogen, oxygen and neon ions are again the strongest emission features in the
low energy part of the spectrum. The high energy region changes noticeably with respect to the
previous case with lower ionization. The strongest Fe~K${\alpha}$ emission in this case is due
to Fe~{\sc xxv} ions at $\sim 6.7$ keV. However, there is a rich structure of emission lines 
with centroid energies that cover from $\sim$ 6.4 keV to 6.7 keV produced by ions at lower ionization
stages (Fe~{\sc xv-xviii}). The Fe~{\sc xxvi} K$\alpha$ emission line is clearly observable at 6.965 keV.
All these lines are superimposed on a very broad, Comptonized profile.
The K${\beta}$ component is replaced by a few distinguishable emission lines mainly due to H- and He-like
iron. As in the previous case, a rich structure of absorption profiles is noticeable between 7.2 and 8.2 keV.
A small absorption edge can be seen just below the RRC located at $\sim 9$ keV.

It is important to notice that the O~{\sc viii} K${\alpha}$ emission line shows the same
kind of broadening as the Fe K line. In fact, the other oxygen lines also show
broadening in some degree, as well as those coming from neon. This is somewhat unexpected,
since photons at those energies would need many scatterings in order to produce such broadening 
if the gas were cold. However, since the temperature is high ($T\sim 10^{6\ \circ}$K) even at 
$\tau \sim 1$, broadening also takes place through the dependence of the Gaussian kernel on the temperature 
(see \S~\ref{secrad}). This also means that the H-like oxygen lines are being produced
efficiently over a large range of optical depths, which can be verified by looking at higher 
ionization parameters. Figure~\ref{sp38} shows the last model with a resolving power
of ${\cal R}\sim 3500$, resulting from a gas at log~$\xi=3.5$ (model 13). Almost no absorption features can be
seen in this spectrum, and the emission comes mostly from recombination of fully ionized ions. Both iron
and oxygen K${\alpha}$ line profiles  are very broad and Comptonized. There are no apparent signs
of broad components in the weaker lines, such as those from neon, although this is likely due to the
fact that the redistribution is so extreme that it completely smears the profile over the 
continuum. Therefore, only those photons produced very close to the surface come out of
the slab unscattered. The emission in the iron K region is exclusively due to H- and He-like
ions. There is strong emission from the Rydberg series of Fe~{\sc xxv}, namely
transitions from 1s 4p, 5p and 6p to the ground state 1s$^2$, at 8.62, 8.83 and 8.94 keV,
respectively.
%
%
\subsection{Anisotropy: incident and viewing angles}\label{seciso}
The boundary condition expressed in Equation~(\ref{eqbc1}) 
depends, in general, on the angle with respect to the normal
on which the radiation is incident. Isotropic illumination implies that the source of
radiation is extended and very close to the region where the calculation takes place. 
On the other hand, locating the source of the external X-rays at a specific point 
far away of the accretion disk constrains the angle at which radiation will penetrate in
the atmosphere. To account for such a situation, the right hand side of Equation~(\ref{eqbc1})
can be expressed as 
\begin{equation}\label{eqbc2}
I_{inc} = I_0 \delta(\mu-\mu_0),
\end{equation}
\begin{figure}
\epsscale{0.9}\plotone{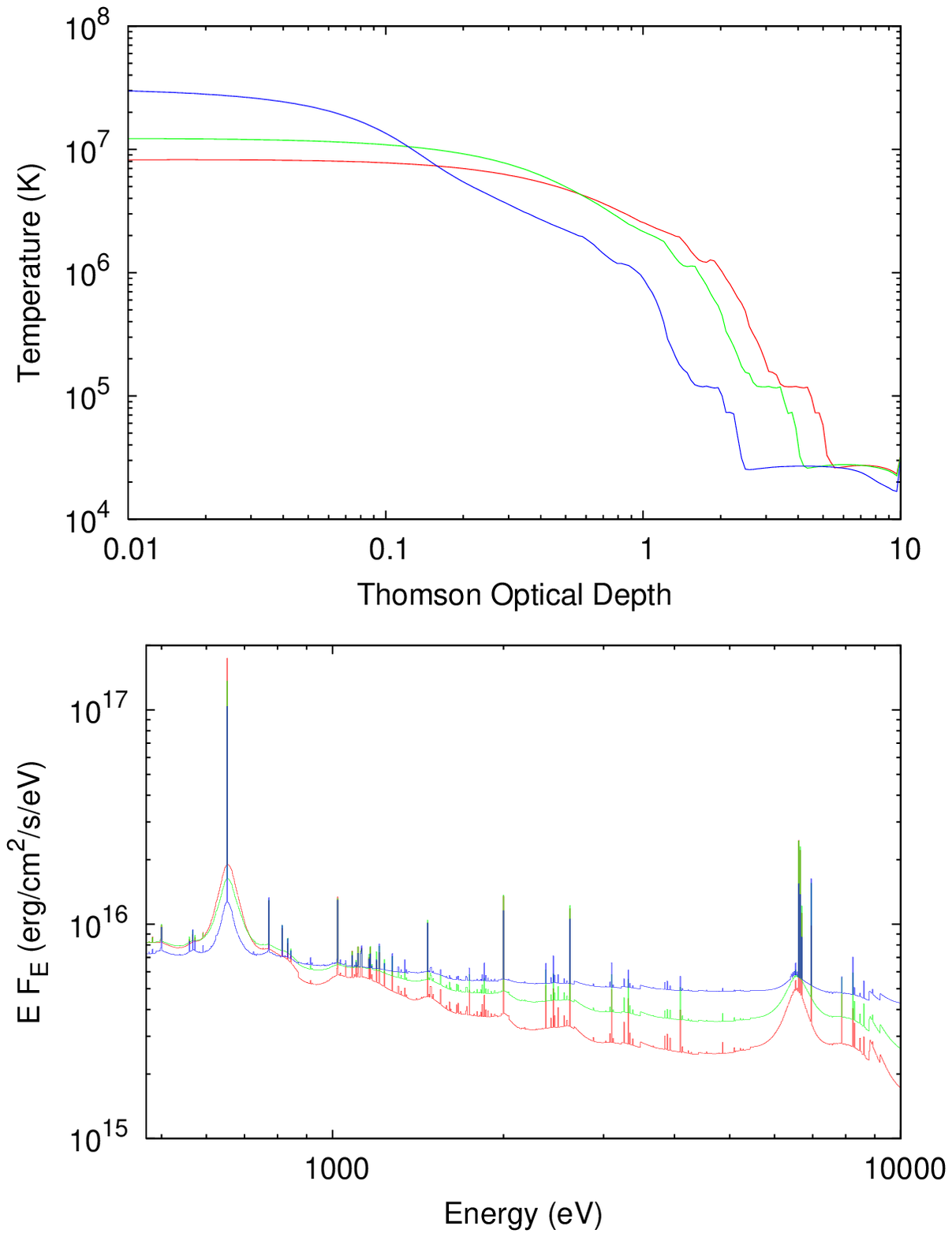}
\caption{Temperature profiles (upper panel) and reflected spectra (lower panel), resulting
from a constant density model with log~$\xi=2.8$ for three different incidence angles (models 14-16).
In the two panels, the {\it red} curves corresponds to $\mu_0=0.95$ ($\theta\approx0^o$, normal incidence);
the {\it green} curves to $\mu_0=0.5$ (or $\theta=60^o$);
and the {\it blue} to $\mu_0=0.05$ ($\theta\approx90^o$, grazing incidence).}
\label{iangle}
\end{figure}
\begin{figure}
\epsscale{1.}\plotone{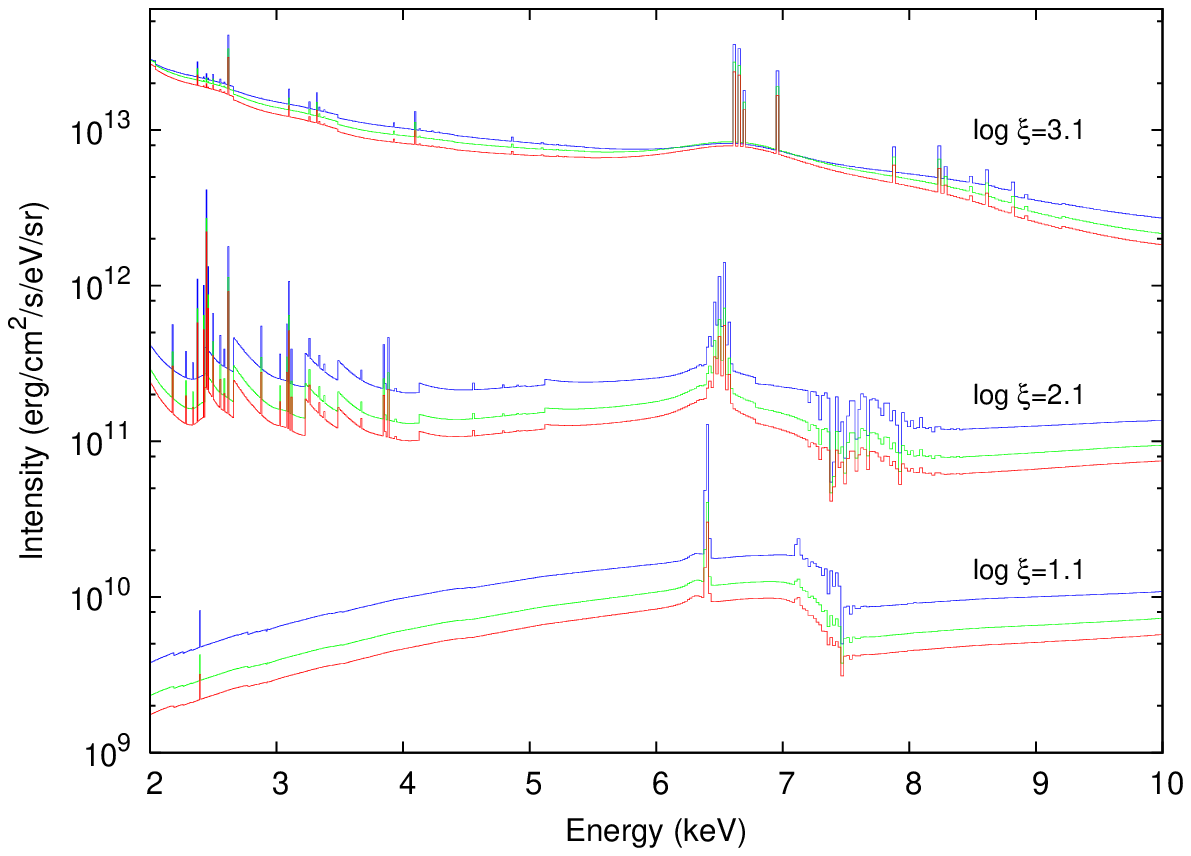}
\caption{Outgoing intensities in the 2-10 keV energy range for
constant density models from 3 different ionization parameters (models 2, 5 and 8),
as viewed at three different angles with respect to the normal. In all cases, the red curves
corresponds to $\mu=0.95$, the green curves to $\mu=0.5$, and the blue curves to $\mu=0.05$.}
\label{vangle}
\end{figure}
where $\mu_0$ is the cosine of the incidence angle. 
The lower panel of Figure~\ref{iangle} shows the reflected spectra for a constant density model in
which log~$\xi=2.8$ for three different incidence angles (models 14-16 in Table~\ref{tamodels}). 
The red curve corresponds to $\mu_0=0.95$ (normal incidence), the green curve
correspond to $\mu_0=0.5$, and the blue curve to $\mu_0=0.05$ (grazing incidence). The upper
panel of the Figure shows the corresponding temperature profiles as a function of the
Thomson optical depth, obtained for each case. At the surface of the disk ($\tau=0$), 
the temperature increases with the incidence angle, such that the grazing incidence shows
the highest temperature of the three cases. However, when the illumination occurs at normal
incidence, the radiation ionizes deeper regions in the slab and there is a larger amount of
hot material for those cases. In this model, the physical quantity that describes the
amount of illumination is the net X-ray flux $F_{\rm x}$, in units of energy per
unit area per unit of time. Since the flux is defined as the second moment of the intensity,
(Equation~\ref{eqhmom}), therefore $F_{\rm x} = \frac{1}{2}\int{I_{inc}\mu d\mu}$, and using
Equation~(\ref{eqbc2})
\begin{equation}\label{bc3}
I_{inc}=\frac{2 F_{\rm x}}{\mu_0} \delta(\mu-\mu_0),
\end{equation}
which means that for the same value of $F_{\rm x}$ (or the same ionization parameter $\xi$),
varying the incidence angle $\mu_0$ effectively varies the intensity of the radiation incident
at the surface. This will, of course, produce more heating and raise the temperature and 
ionization of the top layers in the disk. However, the angular dependence of the radiation
field ensures that in the cases of normal incidence the illuminating radiation reaches deeper
regions in the atmosphere than those of grazing angles, as is expected. Although the effects
on the reflected spectra are not obvious, the spectrum for the grazing incidence
(blue curve) is stronger than the other two, mimicking a case of a higher illumination 
(note that none of the curves have been offset). Nevertheless, the emission lines are, in 
general, very similar in all cases. It is also interesting to see how the temperature 
profiles of the three models converge at large optical depths, since the radiation fields
become more isotropic after many scatterings.

The anisotropy of the reflected radiation field can also be investigated by looking its
angular distribution through different viewing angles. Figure~\ref{vangle} contains the
reflected spectra for 3 different ionization parameters (indicated next to each case), as
is observed from 3 different angles with respect to the normal, in the 2-10 keV energy
range. For consistency the colors
represent the same angles as in Figure~\ref{iangle}, i.e., the red curves indicate $\mu=0.95$
(face-on), the green curves $\mu=0.5$, while the blue curves correspond to $\mu=0.05$ (edge-on).
Note that no rescaling or normalization is applied to these curves, therefore the differences are exclusively 
due to the differences in the ionizing fluxes and on the viewing angles. Moreover, the physical
quantity plotted here is the outgoing specific intensity $I(0,+\mu,E)$ instead of the flux
(as the other Figures), which is an angle averaged quantity. In fact, these 3 cases are
those presented in Figures~\ref{spec} and \ref{zoom} (models 2, 5 and 8 in 
Table~\ref{tamodels}), and thus their temperature profiles correspond to those shown
in Figure~\ref{temp}. In general, the reprocessed features are
stronger when the disk is observed face-on than when the viewing angle is parallel to
the surface. \cite{nay00} found the same tendency in their hydrostatic models, which 
they explained to be a consequence of an effective Thomson depth that changes when the
slab is observed at different angles, according to the geometrical projection $\tau_{eff}=\tau/\mu$.
However, their results suggest that the reflection features almost disappear for small
values of $\mu$, and that no visible iron could been detected in such cases. The results
shown in Figure~\ref{vangle} do not completely agree with this, since the Fe K lines
are strong even for the high ionization case. Nevertheless, one must take this comparison
with care, since all the results presented in this paper correspond to constant density
models, while \cite{nay00} calculations were done under hydrostatic equilibrium. 
In any case, the trend in the emerging spectra shown in Figures~\ref{iangle} and
\ref{vangle} is similar to previous models in the literature, such as those by
\cite{nay00} and \cite{bal01}.
%

\subsection{Iron abundance}\label{seciro}
\begin{figure}
\epsscale{0.8}\plotone{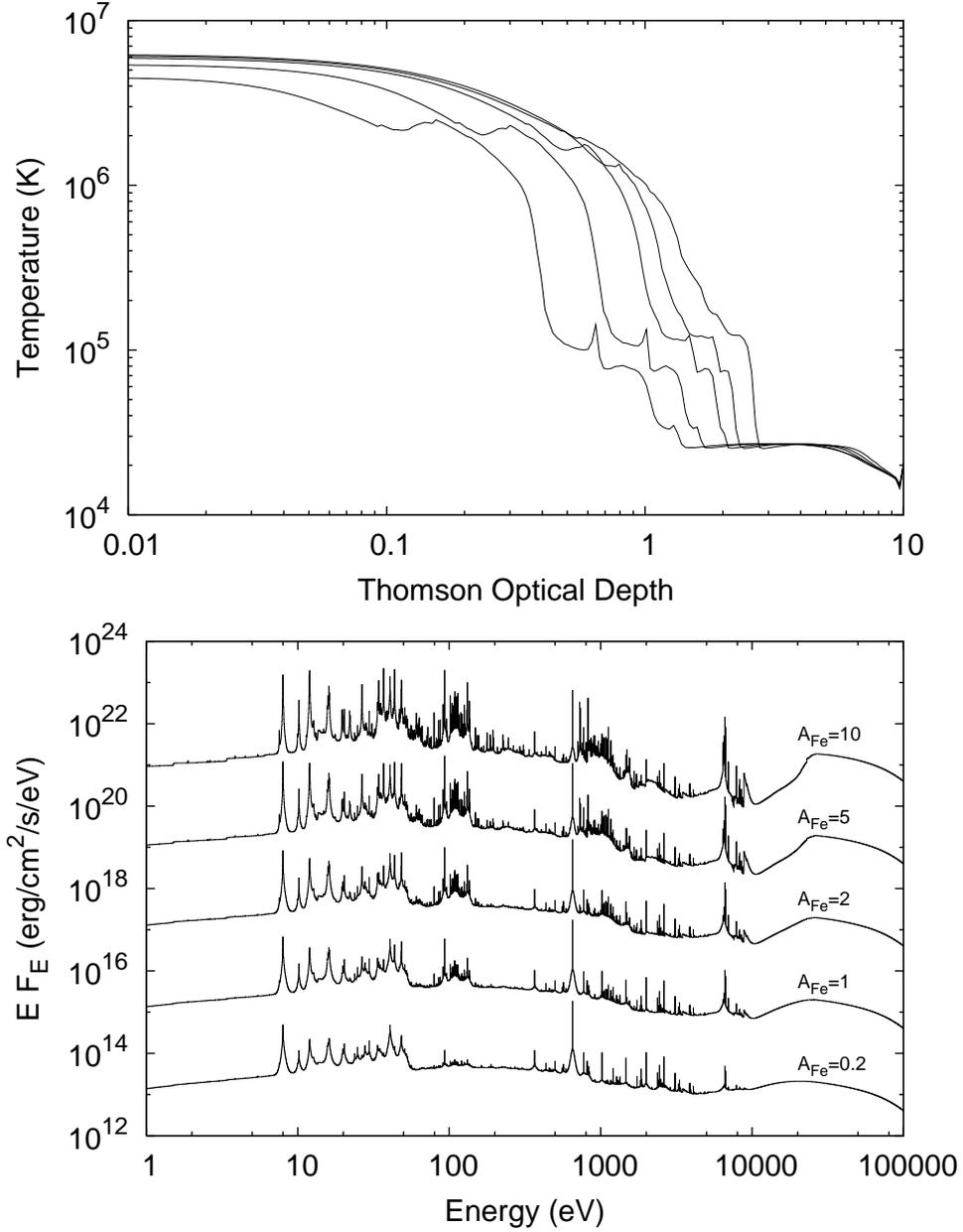}
\caption{{\it Upper panel}: temperature profiles from a constant density model with log~$\xi=2.5$
and different iron abundances with respect to the solar values. From right to left, each curve
correspond to: A$_{\rm Fe}$=0.2, 1, 2, 5 and 10 (models 17, 6 and 18-20, respectively).
{\it Lower panel}: reflected spectra for the same models. The curves are shifted by arbitrary
factors for clarity. These are, from bottom to top: $10^{-2}, 10^0, 10^2, 10^4$, and $10^6$.
The values of the iron abundance are shown at the top of each curve.}
\label{aspec}
\end{figure}
We have also studied the effects of the iron abundance on the ionization of the
atmosphere and the reprocessed spectra. The upper panel of Figure~\ref{aspec} shows the 
temperature profiles along the vertical direction for a gas at log~$\xi=2.5$ when the
iron abundance is assumed to be 0.2, 1, 2, 5 and 10 times its solar value (models 17,
6 and 18-20, respectively). The solar atomic abundances used in all our
calculations are those from \cite{gre96}, and in particular for iron is $3.16\times10^{-5}$
(with respect to the hydrogen).  The curves towards the left of the plot correspond 
to higher values of the iron abundance, which means that the gas is effectively cooler 
than the solar and sub-solar cases. This shows that the iron produces a net
cooling as its abundance is increased. The transition between the hot and cold regions
in the gas occurs at lower optical depths for the super-solar abundance models, 
decreasing the thickness of the hot skin. However, all the models converge to the
same temperature in the cold region of the disk ($T\sim 2.5\times 10^{4\ \circ}$K).

The corresponding reflected spectra are shown in the lower panel of Figure~\ref{aspec}. 
These curves are shifted by arbitrary factors for clarity, which are, from bottom to top:
$10^{-2}, 10^0, 10^2, 10^4$, and $10^6$. The iron emission is enhanced when the abundance is increased, 
as can be clearly seen at $\sim 6$ keV, $\sim 1$ keV and $\sim 0.1$ keV for the K-,
L- and M-shell transitions, respectively. It is also evident that the continuum is 
highly modified in the 1-40 keV energy region, which is where the larger opacity takes
place. However, there are no significant modifications at higher energies (i.e., if these
spectra are placed on the top of each other, there are no differences in the region 
above 40 keV).

The suppression of the continuum combined with the enhancement of the K-shell 
emission lines as the iron abundance is increased shows a significant impact
on the line equivalent width. When solar abundance is assumed, we found an
equivalent width for the Fe K$\alpha$ of $\sim 700$ eV. This value is increased
to about 1.1, 2.3 and 3.9 keV when iron is chosen to be twice, five and ten times
more abundant, respectively. Conversely, this value decreases to $\sim 203$ eV 
for the model when A$_{\rm Fe}=0.2 $.
%
%
\subsection{Comparison with previous models}\label{seccom}
\begin{figure}
\epsscale{0.7}\plotone{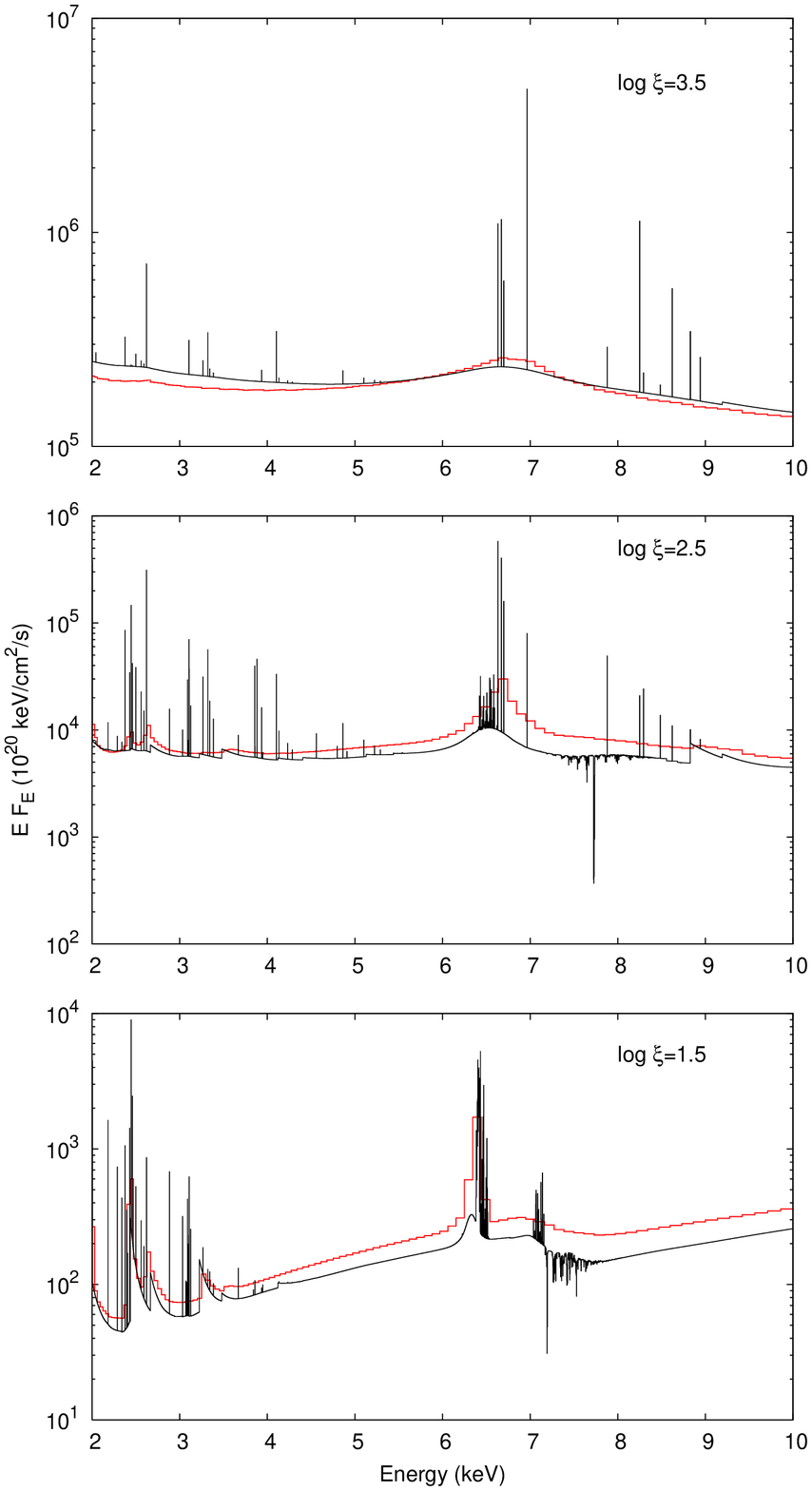}
\caption{Reflected spectra from constant density models for three different
ionization parameters (models 11-13), plotted in the 2-10 keV energy range.
The red curves are the models from {\sc reflion} \citep{ros05}, for the same
input parameters.}
\label{hrcomp}
\end{figure}
Taking into account that our main contribution to the already existing models
of accretion disk is the inclusion of the most complete atomic data available
(particularly important for iron), we compare our results for constant density
accretion disks models with those included in {\sc reflion} \citep{ros05}. 
Such a comparison is shown in Figure~\ref{hrcomp} where the reflected spectra
for three different ionization parameters are shown (log~$\xi=1.5, 2.5$ and 3.5).
In the figure, the black curves correspond to our calculations (models 11-13),
while the red curves are the models from {\sc reflion} with the same input 
parameters.
It is clear from this comparison that these two models
show important differences. The Fe K line profile around 6.4 keV from our model 
is shown to be much more intense but with similar widths, and the absorption edge 
at $\sim 7.5$~keV to be quite strong (especially for the lower ionization). 
In fact, the equivalent widths of the iron K$\alpha$ emission line are
similar in both models for log~$\xi=1.5$, our models predict a value of 610~eV
while {\sc reflion} predicts 793~eV. However, there are
larger discrepancies in the higher ionization cases: our models predict
equivalent widths of 695~eV and 92~eV for log~$\xi=2.5$ and 3.5, respectively;
while {\sc reflion} models show values of 473~eV and 49~eV for the same cases,
respectively. As before, these equivalent widths were calculated using 
a local continuum which is defined by interpolating a straight line between 
two points. In this case the integration was performed between 6 and 7 keV
in order to take into account only the contribution from the K$\alpha$ emission
(since any Fe K$\beta$ emission will occur at energies above 7 keV). Furthermore,
the same integration region is applied to all the models in order to provide
a consistent quantity to compare models with different ionization parameters.

However, we have also calculated the equivalent widths when the
integration region is modified for each model such that the entire broadened line
is included. The new integration regions are: 6.1-6.6~keV, 5.9-7.3~keV, and 5.2-7.8~keV
for models with log~$\xi=1.5, 2.5$ and $3.5$, respectively. The resulting equivalent
withs from our models are 569~eV, 829~eV and 255~eV; while {\sc reflion} predicts
700~eV, 813~eV and 420~eV for the same models, respectively. Clearly, there is
a better agreement when the entire line profile is taken into account, especially
for the two lowest ionization cases. 
Although both set of models share the
same input parameters, in {\sc reflion} the illuminating spectrum extends up to
1 MeV with an exponential cutoff placed at 300 keV, while our models cover up to
210 keV, with the exponential cutoff placed at 200 keV. This could be responsible
for some of the differences seen here, in particular by altering the broad component
of the Fe K emission which is highly affected by the Comptonization of the photons
in the gas. In general, our models predicts stronger Fe K$\alpha$ emission, a more complex 
structure for the Fe absorption edge, and the presence of the Fe K$\beta$ 
emission line at $\sim 7.2$ keV, not seen in the {\sc reflion} models.

Additionally, we have also compared our models with those presented by \cite{nay00}.
In many aspects these models are the most similar to the ones presented in this paper, 
in particular since they have also used the code {\sc xstar} to solve the ionization
structure of the gas. 
By means of visual comparison of their figures we found a good general agreement in the five 
constant density models included in \cite{nay00}. However, our models predicts higher
temperatures in the regions close to the surface of the disk (at small optical 
depths), and lower temperatures at large optical depths; which affects the ionization 
of the gas and the final reprocessed spectrum. This is likely due to the fact that 
their models only explicitly treat a smaller optical depth ($\tau_{\mathrm{Thomson}}=4$)
than ours, and the radiation transmitted through this region is treated approximately.
%
%
\section{Conclusions}\label{seccon}
Answers to outstanding questions concerning accretion disks will not come alone,
but with the need to reproduce and predict new observations and to identify 
spectral profiles at high resolution. In this paper we have presented new models 
for the structure of X-ray illuminated accretion disks 
and their reflected spectra, assuming constant density along the vertical direction.
These models include the most recent and complete atomic data for the iron isonuclear
sequence. The energy resolution used in the reflected spectra exceeds other models
previously published as well as the resolution of the detectors on
current X-ray observatories ({\it Chandra}, {\it XMM Newton, Suzaku}), 
and it is comparable to the expected resolving power of the forthcoming 
International X-Ray Observatory (IXO).

In models with intermediate values of the ionization parameter ($1.5 \leq \rm{log}\xi\leq 3.1$),
the structure of the gas often displays a two temperature regime: a hot skin 
($T > 10^{6\ \circ}$K) close to the surface where the Compton heating and cooling
dominates, and a cold region ($T < 10^{5\ \circ}$K) where the photoelectric 
opacity quickly thermalizes the radiation fields. The thickness of the hot
skin increases with the illumination, and although these solutions are thermally
stable the transition between the hot and cold regions can be sudden. The emerging
spectra corresponding to these models show a clear combination of absorption and 
emission features, particularly for log~$\xi < 2.5$. For higher illumination cases
the effects of Compton scattering become more evident, even in the low-energy
part of the spectrum ($E < 100$ eV). In the high energy part of the spectrum
Compton scattering partially smears the profile of the iron K line. The Compton hump above 10 keV 
is clearly visible in all the models considered here. 

The equivalent width of the Fe K$\alpha$ emission line varies between $\sim$ 400-800
eV for models with $1\lesssim \rm{log}\xi\lesssim 3$. Once the ionization parameter
is larger than $10^3$, the equivalent width decreases drastically. The lowest
value of the equivalent width is $\sim 20$ eV, corresponding to the highest 
illumination case (log~$\xi=3.8$). This behavior resembles the X-ray Baldwin
effect, recently observed in several AGNs. A comparison with other models previously
published such as {\sc reflion} shown important differences in the structure of
the iron K lines. In particular, the equivalent widths in those models can
differ by a factor of 2 when compared to the ones shown in this paper.

In general, our simulations show that the K-shell atomic data is
crucial to properly model the structure and profile of the iron lines.
These results also suggest that the line emission from Fe ions in different ionization stages
and Comptonization of high energy photons by cold electrons can be responsible for
significant line broadening. These processes need to be taken into account since
they can be mistaken for relativistic effects, especially in cases when the gas is
partially or high ionized.

Because the state of the gas in the accretion disk depends on the radiation field
at each point, and the ionization balance required for a realistic definition of
the source function in the radiation transfer depends on both the temperature and
density of the gas, a self-consistent approach must be taken into account in order
to solve the hydrostatic equilibrium equation instead of assuming constant density.
This will be considered as the next step of this project. Future work will also 
consider the inclusion of new K-shell atomic data recently calculated for the Ne,
Mg, Si, S, Ar and Ca by \cite{pal08a} and \cite{wit09}, as well as for the
nickel \citep{pal08b}, and nitrogen \citep{gar09a} isonuclear sequences.
%
\acknowledgments
We would like to thank the referee Prof. Randy Ross for his valuable comments on
the paper. This work was supported by a grant from the NASA astrophysics theory program 
05-ATP05-18. This research has made use of NASA's Astrophysics Data System.
%
%
\bibliographystyle{apj}
\bibliography{my-references}

\begin{thebibliography}{76}
\expandafter\ifx\csname natexlab\endcsname\relax\def\natexlab#1{#1}\fi

\bibitem[{{Badnell}(1997)}]{bad97}
{Badnell}, N.~R. 1997, J. Phys. B: At. Mol. Opt. Phys., 30, 1

\bibitem[{{Ballantyne} {et~al.}(2001){Ballantyne}, {Ross}, \& {Fabian}}]{bal01}
{Ballantyne}, D.~R., {Ross}, R.~R., \& {Fabian}, A.~C. 2001, \mnras, 327, 10

\bibitem[{{Bautista} \& {Kallman}(2001)}]{bau01}
{Bautista}, M.~A., \& {Kallman}, T.~R. 2001, \apjs, 134, 139

\bibitem[{{Bautista} {et~al.}(2003){Bautista}, {Mendoza}, {Kallman}, \&
  {Palmeri}}]{bau03}
{Bautista}, M.~A., {Mendoza}, C., {Kallman}, T.~R., \& {Palmeri}, P. 2003,
  \aap, 403, 339

\bibitem[{{Bautista} {et~al.}(2004){Bautista}, {Mendoza}, {Kallman}, \&
  {Palmeri}}]{bau04}
---. 2004, \aap, 418, 1171

\bibitem[{{Berrington} {et~al.}(1987){Berrington}, {Burke}, {Butler}, {Seaton},
  {Storey}, {Taylor}, \& {Yan}}]{ber87}
{Berrington}, K.~A., {Burke}, P.~G., {Butler}, K., {Seaton}, M.~J., {Storey},
  P.~J., {Taylor}, K.~T., \& {Yan}, Y. 1987, J. Phys. B: At. Mol. Opt. Phys.,
  20, 6379

\bibitem[{{Bhattacharyya} \& {Strohmayer}(2007)}]{bha07}
{Bhattacharyya}, S., \& {Strohmayer}, T.~E. 2007, \apjl, 664, L103

\bibitem[{{Blackman}(1999)}]{bla99}
{Blackman}, E.~G. 1999, \mnras, 306, L25

\bibitem[{{Brenneman} \& {Reynolds}(2006)}]{bre06}
{Brenneman}, L.~W., \& {Reynolds}, C.~S. 2006, \apj, 652, 1028

\bibitem[{{Buff} \& {McCray}(1974)}]{buf74}
{Buff}, J., \& {McCray}, R. 1974, \apj, 189, 147

\bibitem[{{Cackett} {et~al.}(2009){Cackett}, {Altamirano}, {Patruno}, {Miller},
  {Reynolds}, {Linares}, \& {Wijnands}}]{cac09}
{Cackett}, E.~M., {Altamirano}, D., {Patruno}, A., {Miller}, J.~M., {Reynolds},
  M., {Linares}, M., \& {Wijnands}, R. 2009, \apjl, 694, L21

\bibitem[{{Cackett} {et~al.}(2008){Cackett}, {Miller}, {Bhattacharyya},
  {Grindlay}, {Homan}, {van der Klis}, {Miller}, {Strohmayer}, \&
  {Wijnands}}]{cac08}
{Cackett}, E.~M., {et~al.} 2008, \apj, 674, 415

\bibitem[{{Chandrasekhar}(1960)}]{cha60}
{Chandrasekhar}, S. 1960, {Radiative transfer} (New York: Dover)

\bibitem[{{Cowan}(1981)}]{cow81}
{Cowan}, R.~D. 1981, {The theory of atomic structure and spectra} (Berkeley,
  CA: Univ. of California Press)

\bibitem[{{Cunto} {et~al.}(1993){Cunto}, {Mendoza}, {Ochsenbein}, \&
  {Zeippen}}]{cun93}
{Cunto}, W., {Mendoza}, C., {Ochsenbein}, F., \& {Zeippen}, C.~J. 1993, \aap,
  275, L5+

\bibitem[{{Czerny} \& {Zycki}(1994)}]{cze94}
{Czerny}, B., \& {Zycki}, P.~T. 1994, \apjl, 431, L5

\bibitem[{{Done} {et~al.}(1992){Done}, {Mulchaey}, {Mushotzky}, \&
  {Arnaud}}]{don92}
{Done}, C., {Mulchaey}, J.~S., {Mushotzky}, R.~F., \& {Arnaud}, K.~A. 1992,
  \apj, 395, 275

\bibitem[{{Dumont} {et~al.}(2002){Dumont}, {Czerny}, {Collin}, \&
  {Zycki}}]{dum02}
{Dumont}, A.-M., {Czerny}, B., {Collin}, S., \& {Zycki}, P.~T. 2002, \aap, 387,
  63

\bibitem[{{Fabian} {et~al.}(1989){Fabian}, {Rees}, {Stella}, \&
  {White}}]{fab89}
{Fabian}, A.~C., {Rees}, M.~J., {Stella}, L., \& {White}, N.~E. 1989, \mnras,
  238, 729

\bibitem[{{Garc{\'{\i}}a} {et~al.}(2005){Garc{\'{\i}}a}, {Mendoza}, {Bautista},
  {Gorczyca}, {Kallman}, \& {Palmeri}}]{gar05}
{Garc{\'{\i}}a}, J., {Mendoza}, C., {Bautista}, M.~A., {Gorczyca}, T.~W.,
  {Kallman}, T.~R., \& {Palmeri}, P. 2005, \apjs, 158, 68

\bibitem[{{Garc{\'{\i}}a} {et~al.}(2009){Garc{\'{\i}}a}, {Kallman},
  {Witthoeft}, {Behar}, {Mendoza}, {Palmeri}, {Quinet}, {Bautista}, \&
  {Klapisch}}]{gar09a}
{Garc{\'{\i}}a}, J., {et~al.} 2009, \apjs, 185, 477

\bibitem[{{George} \& {Fabian}(1991)}]{geo91}
{George}, I.~M., \& {Fabian}, A.~C. 1991, \mnras, 249, 352

\bibitem[{{Gorczyca} \& {McLaughlin}(2005)}]{gor05}
{Gorczyca}, T., \& {McLaughlin}, B. 2005, APS Meeting Abstracts, D6037+

\bibitem[{{Gottwald} {et~al.}(1995){Gottwald}, {Parmar}, {Reynolds}, {White},
  {Peacock}, \& {Taylor}}]{got95}
{Gottwald}, M., {Parmar}, A.~N., {Reynolds}, A.~P., {White}, N.~E., {Peacock},
  A., \& {Taylor}, B.~G. 1995, \aaps, 109, 9

\bibitem[{{Grevesse} {et~al.}(1996){Grevesse}, {Noels}, \& {Sauval}}]{gre96}
{Grevesse}, N., {Noels}, A., \& {Sauval}, A.~J. 1996, in Astronomical Society
  of the Pacific Conference Series, Vol.~99, Cosmic Abundances, ed. S.~S.
  {Holt} \& G.~{Sonneborn}, 117--+

\bibitem[{{Hummer} {et~al.}(1993){Hummer}, {Berrington}, {Eissner}, {Pradhan},
  {Saraph}, \& {Tully}}]{hum93}
{Hummer}, D.~G., {Berrington}, K.~A., {Eissner}, W., {Pradhan}, A.~K.,
  {Saraph}, H.~E., \& {Tully}, J.~A. 1993, \aap, 279, 298

\bibitem[{{Illarionov} {et~al.}(1979){Illarionov}, {Kallman}, {McCray}, \&
  {Ross}}]{ill79}
{Illarionov}, A., {Kallman}, T., {McCray}, R., \& {Ross}, R. 1979, \apj, 228,
  279

\bibitem[{{Iwasawa} {et~al.}(1999){Iwasawa}, {Fabian}, {Young}, {Inoue}, \&
  {Matsumoto}}]{iwa99}
{Iwasawa}, K., {Fabian}, A.~C., {Young}, A.~J., {Inoue}, H., \& {Matsumoto}, C.
  1999, \mnras, 306, L19

\bibitem[{{Iwasawa} \& {Taniguchi}(1993)}]{iwa93}
{Iwasawa}, K., \& {Taniguchi}, Y. 1993, \apjl, 413, L15

\bibitem[{{Kaastra} \& {Mewe}(1993)}]{kas93}
{Kaastra}, J.~S., \& {Mewe}, R. 1993, \aaps, 97, 443

\bibitem[{{Kallman} \& {Bautista}(2001)}]{kal01}
{Kallman}, T., \& {Bautista}, M. 2001, \apjs, 133, 221

\bibitem[{{Kallman} {et~al.}(2004){Kallman}, {Palmeri}, {Bautista}, {Mendoza},
  \& {Krolik}}]{kal04}
{Kallman}, T.~R., {Palmeri}, P., {Bautista}, M.~A., {Mendoza}, C., \& {Krolik},
  J.~H. 2004, \apjs, 155, 675

\bibitem[{{Krolik} {et~al.}(1994){Krolik}, {Madau}, \& {Zycki}}]{kro94}
{Krolik}, J.~H., {Madau}, P., \& {Zycki}, P.~T. 1994, \apjl, 420, L57

\bibitem[{{Krolik} {et~al.}(1981){Krolik}, {McKee}, \& {Tarter}}]{kro81}
{Krolik}, J.~H., {McKee}, C.~F., \& {Tarter}, C.~B. 1981, \apj, 249, 422

\bibitem[{{Landi} \& {Phillips}(2006)}]{lan06}
{Landi}, E., \& {Phillips}, K.~J.~H. 2006, \apjs, 166, 421

\bibitem[{{Laor}(1991)}]{lao91}
{Laor}, A. 1991, \apj, 376, 90

\bibitem[{{Lightman} {et~al.}(1981){Lightman}, {Lamb}, \& {Rybicki}}]{lig81}
{Lightman}, A.~P., {Lamb}, D.~Q., \& {Rybicki}, G.~B. 1981, \apj, 248, 738

\bibitem[{{Lightman} \& {Rybicki}(1980)}]{lig80}
{Lightman}, A.~P., \& {Rybicki}, G.~B. 1980, \apj, 236, 928

\bibitem[{{Lightman} \& {White}(1988)}]{lig88}
{Lightman}, A.~P., \& {White}, T.~R. 1988, \apj, 335, 57

\bibitem[{{Magdziarz} \& {Zdziarski}(1995)}]{mag95}
{Magdziarz}, P., \& {Zdziarski}, A.~A. 1995, \mnras, 273, 837

\bibitem[{{Markowitz} {et~al.}(2007){Markowitz}, {Takahashi}, {Watanabe},
  {Nakazawa}, {Fukazawa}, {Kokubun}, {Makishima}, {Awaki}, {Bamba}, {Isobe},
  {Kataoka}, {Madejski}, {Mushotzky}, {Okajima}, {Ptak}, {Reeves}, {Ueda},
  {Yamasaki}, \& {Yaqoob}}]{mar07}
{Markowitz}, A., {et~al.} 2007, \apj, 665, 209

\bibitem[{{Matt} {et~al.}(1993){Matt}, {Fabian}, \& {Ross}}]{mat93}
{Matt}, G., {Fabian}, A.~C., \& {Ross}, R.~R. 1993, \mnras, 262, 179

\bibitem[{{Matt} {et~al.}(1996){Matt}, {Fabian}, \& {Ross}}]{mat96}
---. 1996, \mnras, 278, 1111

\bibitem[{{Matt} {et~al.}(1991){Matt}, {Perola}, \& {Piro}}]{mat91}
{Matt}, G., {Perola}, G.~C., \& {Piro}, L. 1991, \aap, 247, 25

\bibitem[{{Mendoza} {et~al.}(2004){Mendoza}, {Kallman}, {Bautista}, \&
  {Palmeri}}]{men04}
{Mendoza}, C., {Kallman}, T.~R., {Bautista}, M.~A., \& {Palmeri}, P. 2004,
  \aap, 414, 377

\bibitem[{{Mihalas}(1978)}]{mih78}
{Mihalas}, D. 1978, {Stellar atmospheres} (2nd ed.; San Francisco, CA: Freeman)

\bibitem[{{Nayakshin} \& {Kallman}(2001)}]{nay01}
{Nayakshin}, S., \& {Kallman}, T.~R. 2001, \apj, 546, 406

\bibitem[{{Nayakshin} {et~al.}(2000){Nayakshin}, {Kazanas}, \&
  {Kallman}}]{nay00}
{Nayakshin}, S., {Kazanas}, D., \& {Kallman}, T.~R. 2000, \apj, 537, 833

\bibitem[{{Noble} {et~al.}(2010){Noble}, {Krolik}, \& {Hawley}}]{nob10}
{Noble}, S.~C., {Krolik}, J.~H., \& {Hawley}, J.~F. 2010, ArXiv e-prints

\bibitem[{{Page} {et~al.}(2004){Page}, {O'Brien}, {Reeves}, \&
  {Turner}}]{pag04}
{Page}, K.~L., {O'Brien}, P.~T., {Reeves}, J.~N., \& {Turner}, M.~J.~L. 2004,
  \mnras, 347, 316

\bibitem[{{Palmeri} {et~al.}(2002){Palmeri}, {Mendoza}, {Kallman}, \&
  {Bautista}}]{pal02}
{Palmeri}, P., {Mendoza}, C., {Kallman}, T.~R., \& {Bautista}, M.~A. 2002,
  \apjl, 577, L119

\bibitem[{{Palmeri} {et~al.}(2003{\natexlab{a}}){Palmeri}, {Mendoza},
  {Kallman}, \& {Bautista}}]{pal03a}
---. 2003{\natexlab{a}}, \aap, 403, 1175

\bibitem[{{Palmeri} {et~al.}(2003{\natexlab{b}}){Palmeri}, {Mendoza},
  {Kallman}, {Bautista}, \& {Mel{\'e}ndez}}]{pal03b}
{Palmeri}, P., {Mendoza}, C., {Kallman}, T.~R., {Bautista}, M.~A., \&
  {Mel{\'e}ndez}, M. 2003{\natexlab{b}}, \aap, 410, 359

\bibitem[{{Palmeri} {et~al.}(2008{\natexlab{a}}){Palmeri}, {Quinet}, {Mendoza},
  {Bautista}, {Garc{\'{\i}}a}, \& {Kallman}}]{pal08a}
{Palmeri}, P., {Quinet}, P., {Mendoza}, C., {Bautista}, M.~A., {Garc{\'{\i}}a},
  J., \& {Kallman}, T.~R. 2008{\natexlab{a}}, \apjs, 177, 408

\bibitem[{{Palmeri} {et~al.}(2008{\natexlab{b}}){Palmeri}, {Quinet}, {Mendoza},
  {Bautista}, {Garc{\'{\i}}a}, {Witthoeft}, \& {Kallman}}]{pal08b}
{Palmeri}, P., {Quinet}, P., {Mendoza}, C., {Bautista}, M.~A., {Garc{\'{\i}}a},
  J., {Witthoeft}, M.~C., \& {Kallman}, T.~R. 2008{\natexlab{b}}, \apjs, 179,
  542

\bibitem[{{Poutanen} {et~al.}(1996){Poutanen}, {Nagendra}, \&
  {Svensson}}]{pou96}
{Poutanen}, J., {Nagendra}, K.~N., \& {Svensson}, R. 1996, \mnras, 283, 892

\bibitem[{{Pozdniakov} {et~al.}(1979){Pozdniakov}, {Sobol}, \&
  {Sunyaev}}]{poz79}
{Pozdniakov}, L.~A., {Sobol}, I.~M., \& {Sunyaev}, R.~A. 1979, \aap, 75, 214

\bibitem[{{Ralchenko} {et~al.}(2008){Ralchenko}, {Kramida}, {Reader}, \& {NIST
  ADS Team}}]{ral08}
{Ralchenko}, Y., {Kramida}, A.~E., {Reader}, J., \& {NIST ADS Team}. 2008, NIST
  Atomic Spectra Database, version 3.1.5 (Gaithersburg: NIST),
  http://physics.nist.gov/asd3

\bibitem[{{Reeves} {et~al.}(2007){Reeves}, {Awaki}, {Dewangan}, {Fabian},
  {Fukazawa}, {Gallo}, {Griffiths}, {Inoue}, {Kunieda}, {Markowitz},
  {Miniutti}, {Mizuno}, {Mushotzky}, {Okajima}, {Ptak}, {Takahashi},
  {Terashima}, {Ushio}, {Watanabe}, {Yamasaki}, {Yamauchi}, \&
  {Yaqoob}}]{ree07}
{Reeves}, J.~N., {et~al.} 2007, \pasj, 59, 301

\bibitem[{{Reis} {et~al.}(2009){Reis}, {Fabian}, \& {Young}}]{rei09}
{Reis}, R.~C., {Fabian}, A.~C., \& {Young}, A.~J. 2009, \mnras, L297+

\bibitem[{{Ross} \& {Fabian}(1993)}]{ros93}
{Ross}, R.~R., \& {Fabian}, A.~C. 1993, \mnras, 261, 74

\bibitem[{{Ross} \& {Fabian}(2005)}]{ros05}
---. 2005, \mnras, 358, 211

\bibitem[{{Ross} \& {Fabian}(2007)}]{ros07}
---. 2007, \mnras, 381, 1697

\bibitem[{{Ross} {et~al.}(1996){Ross}, {Fabian}, \& {Brandt}}]{ros96}
{Ross}, R.~R., {Fabian}, A.~C., \& {Brandt}, W.~N. 1996, \mnras, 278, 1082

\bibitem[{{Ross} {et~al.}(1978){Ross}, {Weaver}, \& {McCray}}]{ros78}
{Ross}, R.~R., {Weaver}, R., \& {McCray}, R. 1978, \apj, 219, 292

\bibitem[{{Rozanska} \& {Czerny}(1996)}]{roz96}
{Rozanska}, A., \& {Czerny}, B. 1996, Acta Astron., 46, 233

\bibitem[{{Seaton}(1987)}]{sea87}
{Seaton}, M. 1987, J. Phys. B: At. Mol. Opt. Phys., 20, 6363

\bibitem[{{Shakura} \& {Sunyaev}(1973)}]{sak73}
{Shakura}, N.~I., \& {Sunyaev}, R.~A. 1973, \aap, 24, 337

\bibitem[{{Summers}(2004)}]{sum04}
{Summers}, H.~P. 2004, The ADAS User Manual, version 2.6,
  http://adas.phys.strath.ac.uk

\bibitem[{{Tanaka} {et~al.}(1995){Tanaka}, {Nandra}, {Fabian}, {Inoue},
  {Otani}, {Dotani}, {Hayashida}, {Iwasawa}, {Kii}, {Kunieda}, {Makino}, \&
  {Matsuoka}}]{tan95}
{Tanaka}, Y., {et~al.} 1995, \nat, 375, 659

\bibitem[{{Tarter} {et~al.}(1969){Tarter}, {Tucker}, \& {Salpeter}}]{tar69}
{Tarter}, C.~B., {Tucker}, W.~H., \& {Salpeter}, E.~E. 1969, \apj, 156, 943

\bibitem[{{Verner} \& {Yakovlev}(1995)}]{ver95}
{Verner}, D.~A., \& {Yakovlev}, D.~G. 1995, \aaps, 109, 125

\bibitem[{{Watanabe} {et~al.}(2003){Watanabe}, {Sako}, {Ishida}, {Ishisaki},
  {Kahn}, {Kohmura}, {Morita}, {Nagase}, {Paerels}, \& {Takahashi}}]{wat03}
{Watanabe}, S., {et~al.} 2003, \apjl, 597, L37

\bibitem[{{Witthoeft} {et~al.}(2009){Witthoeft}, {Bautista}, {Mendoza},
  {Kallman}, {Palmeri}, \& {Quinet}}]{wit09}
{Witthoeft}, M.~C., {Bautista}, M.~A., {Mendoza}, C., {Kallman}, T.~R.,
  {Palmeri}, P., \& {Quinet}, P. 2009, \apjs, 182, 127

\bibitem[{{Yaqoob} {et~al.}(2007){Yaqoob}, {Murphy}, {Griffiths}, {Haba},
  {Inoue}, {Itoh}, {Kelley}, {Kokubun}, {Markowitz}, {Mushotzky}, {Okajima},
  {Ptak}, {Reeves}, {Serlemitsos}, {Takahashi}, \& {Terashima}}]{yaq07}
{Yaqoob}, T., {et~al.} 2007, \pasj, 59, 283

\bibitem[{{Zycki} {et~al.}(1994){Zycki}, {Krolik}, {Zdziarski}, \&
  {Kallman}}]{zyc94}
{Zycki}, P.~T., {Krolik}, J.~H., {Zdziarski}, A.~A., \& {Kallman}, T.~R. 1994,
  \apj, 437, 597

\end{thebibliography}
%
\end{document}